\documentclass[runningheads]{llncs}
\usepackage[T1]{fontenc}

\usepackage{multirow}%
\usepackage{amsmath,amssymb,amsfonts}%
\usepackage{booktabs}%
\usepackage{algorithm}%
\usepackage{algorithmicx}%
\usepackage{algpseudocode}
\usepackage[hidelinks,colorlinks=true,linkcolor=blue,citecolor=blue,urlcolor=blue]{hyperref}

\usepackage{graphicx}
\begin{document}

\title{The anonymization problem in social networks}

\author{R. G. de Jong*\inst{1}\orcidID{0000-0002-2680-2816} \and
M. P. J. van der Loo\inst{2,1}\orcidID{
    0000-0002-9807-4686} \and
F. W. Takes\inst{1}\orcidID{  0000-0001-5468-1030}}
\authorrunning{R. G. de Jong, M. P. J. van der Loo and F. W. Takes}
%

\institute{LIACS, Leiden University, Leiden, the Netherlands, 
\and
CBS, Statistics Netherlands, Den Haag, the Netherlands\\
*Corresponding author: r.g.de.jong@liacs.leidenuniv.nl
}

\maketitle              

\begin{abstract}
This paper introduces a unified computational framework for the anonymization problem in social networks, where the objective is to maximize node anonymity through graph alterations.
We define three variants of the underlying optimization problem: full, partial and budgeted anonymization.
In each variant, the objective is to maximize the number of $k$-anonymous nodes, i.e., nodes for which at least $k-1$ other nodes are equivalent under a particular anonymity measure. 
We propose four new heuristic network anonymization algorithms and implement these in ANO-NET, a reusable computational framework.
Experiments on three common graph models and 19 real-world network datasets yield three empirical findings. 
First, regarding the method of alteration, experiments on graph models show that random edge deletion is more effective than edge rewiring and addition. 
Second, we show that the choice of anonymity measure strongly affects both initial network anonymity and the difficulty of anonymization.
This highlights the importance of careful measure selection, matching a realistic attacker scenario.  
Third, comparing the four proposed algorithms and an edge sampling baseline from the literature, we find that an approach which preferentially deletes edges affecting structurally unique nodes, consistently outperforms heuristics based solely on network structure.
Overall, our best performing algorithm retains on average 14 times more edges in full anonymization. 
Moreover, it yields 4.8 times more anonymous nodes than the baseline in the budgeted variant.
On top of that, the best performing algorithm achieves a better trade-off between anonymity and data utility. 
This work provides a foundation for the future development of effective network anonymization algorithms.

    
\end{abstract}

\section{Introduction}\label{sec:intro}
In social network analysis, also known as network science, interactions between people are often the object of study for various applications and downstream tasks, including influence maximization~\cite{chen2009efficient} and understanding epidemic spread~\cite{barabasi2016network}. 
To conduct such research, it is desirable to have large real-world networks of people that are as realistic as possible. 
However, the identity of individuals in these networks can be revealed based on structural network information, even after removing
unique identifiers, i.e., after pseudonymization~\cite{dejong2023effect,romanini2021privacy}.
As a result, sharing or publishing social network data can lead to a breach in the privacy of the people represented.
To solve this problem, sometimes also referred to as identity anonymization~\cite{lu2012fast}, various works introduced methods to share or publish networks in a privacy aware manner, including clustering~\cite{campan2008data}, 
differential privacy~\cite{jiang2021applications} 
and $k$-anonymity~\cite{hay2008resisting,liu2008towards}. 
Each of these approaches anonymizes the data in a different way.
In this work, we focus on $k$-anonymity methods, which, contrary to approaches mentioned above, allow the creation of an altered anonymized version of the original full network dataset, retaining all individuals.  

In $k$-anonymity, the anonymity of a network as a whole is determined as the fraction of anonymous nodes. 
A node is $k$-anonymous if it has, according to a particular \emph{anonymity measure}, the same \emph{signature} as at least $k-1$ other nodes. 
For example, a node is $k$-anonymous with respect to the measure of degree if it has the same degree value as $k-1$ other nodes.
The choice of measure essentially determines the attacker scenario against which one protects~\cite{dejong2024comparison}. 

Much previous work focuses on $k=2$, measuring network anonymity as uniqueness: the fraction of nodes with a unique signature~\cite{dejong2023effect,romanini2021privacy}. 
Given a network and an anonymity measure, the aim of \emph{network anonymization} is then to decrease the uniqueness to a desired level by means of graph alterations, while ensuring that the resulting network data remain suitable for its intended purposes. 
Consequently, network anonymization involves a trade-off between anonymity and \emph{data utility}. 
It is a known fact that altering fewer edges leads to higher utility~\cite{borgatti2006robustness}. 
Utility can also be measured in terms of how well structural network properties are preserved, or as the extent to which performance in network analysis tasks such as community detection is retained after anonymization.

The time complexity of anonymizing a network with as few alterations as possible depends on the chosen anonymity measure.
While for degree, anonymization can be done in quadratic time~\cite{lu2012fast}, the problem is NP-hard when the exact structure of the 1-neighborhood and node labels are considered~\cite{zhou2011k}.
Furthermore, the objective function in the anonymization problem is neither monotonic nor submodular, elegant properties that enabled the development of efficient methods for other common network analysis problems such as influence maximization~\cite{chen2009efficient}. 
Therefore, existing work on the anonymization of networks focuses on (meta)heuristic algorithms, yet always for either a specific anonymity measure~\cite{liu2008towards,zhou2011k}, tailored to a specific type of network~\cite{malik2024towards}, or with a focus on random alterations~\cite{hay2008resisting,romanini2021privacy}.

In this paper, we unite these lines of research and introduce a general formulation of the anonymization problem in the form of three meaningful variants, being 1)~full, 2)~partial and 3)~budgeted anonymization. 
In addition, we consider four commonly used anonymity measures from the literature, providing insights into how the choice of measure affects the anonymization process. 
For the latter task, we propose four new measure-agnostic heuristic anonymization algorithms. 
By means of experiments on both graph models and real-world network data, we independently investigate the effect of A) the chosen alteration operation, B) the anonymity measure, C) the anonymization algorithm and D) the effect of the problem variants on the attained utility. 
Together with this paper, we release ANO-NET\footnote{\url{https://github.com/RacheldeJong/ANONET}},  a computational framework for anonymizing networks that enables researchers to implement and evaluate anonymization algorithms across different anonymity measures and different variants of the problem. The framework is accompanied by a representative real-world benchmark dataset. 

The remainder of this paper is structured as follows.
First, Section~\ref{sec:related} reviews related work.
Section~\ref{sec:prel} covers preliminaries on networks and anonymization.
In Section~\ref{sec:problem}, we formally define the anonymization problem and its variants.
Section~\ref{sec:approach} presents our framework for the anonymization process and the proposed heuristic algorithms.
We discuss experiments and results in Section~\ref{sec:results}.
Finally, Section~\ref{sec:conc} concludes the paper and outlines directions for future work. 

\section{Related work}\label{sec:related}
Below, we discuss the most common measures of anonymity and their commonly used algorithms for anonymization (for more details, see for example~\cite{liu2008towards,zhou2011k}). 

\textbf{Degree.}
The simplest anonymity measure is the node degree, for which exact algorithms minimizing the number of alterations~\cite{liu2008towards} have been devised.

\textbf{Neighborhood.}
Node anonymity measures based on the ego network or $d$-neighborhood capture more structural information~\cite{alavi2019attacker,dejong2023algorithms,romanini2021privacy,zhou2011k}. 
In  node labeled networks, anonymization under this measure is NP-hard~\cite{zhou2011k}, making exact approaches infeasible.
Consequently, a greedy algorithm that modifies node neighborhoods to resemble those of similar nodes~\cite{zhou2011k}, a genetic algorithm~\cite{alavi2019attacker} and an edge sampling method~\cite{romanini2021privacy} were proposed. 

\textbf{Graph invariants.}
The work of~\cite{dejong2024comparison} introduced measures based on graph invariants: the number of nodes and edges in the $d$-neighborhood of a considered node, and the degree distribution within the $d$-neighborhood.

\textbf{VRQ.}
Vertex Refinement Queries (VRQ)~\cite{hay2008resisting} consider the multiset of node degrees within distance $d$ of a given node as the measure of anonymity. Anonymization algorithms include random rewiring and algorithms that find similar neighborhoods~\cite{thompson2009union}.

In this paper, we contribute to the existing literature by formally introducing and exploring three meaningful variants of the anonymization problem. 
Instead of proposing algorithms tailored to a specific measure (i.e., a specific attacker scenario), we present a general framework applicable to all variants, and propose several measure-agnostic heuristic anonymization algorithms.
Crucially, we empirically examine the impact of different anonymity measures on the anonymization process, and analyze the trade-off between anonymity and utility.

\section{Preliminaries}\label{sec:prel}
In this section we formally introduce concepts used throughout this paper, including notions of networks, anonymity and the graph alteration operations.

\subsection{Networks}\label{sub:net}
We define a network as an undirected graph $G = (V, E)$ containing a set of nodes $V$ and a set of edges $\{v, w\} \in E$ between nodes $v, w \in V$.
The degree of a node is the number of connections it has: $degree(v) = |\{w : \{v, w\} \in E \}|$.
The clustering coefficient of 
a node $v$ equals the number of triangles it is part of,
divided by the maximum number of triangles it could form. 
The distance between two nodes, denoted $dist(v, w)$, is the minimal number of edges that needs to be traversed to move from node $v$ to node $w$.
It follows that $dist(v, v) = 0$ and if no path exists, i.e., if the nodes belong to different components, we let $dist(v, w) = \infty$.
The component with the largest number of nodes is called the largest connected component (LCC) or giant component.

The $d$-neighborhood of a node, $N_d(v) = (V_{N_d(v)}, E_{N_d(v)})$, is the subgraph consisting of all nodes that are at most at distance $d$ of node $v$, and all edges between them.
Two $d$-neighborhoods are structurally indistinguishable if they are \emph{isomorphic}. 
We can determine whether two $d$-neighborhoods are isomorphic by comparing their \emph{canonical labeling}~\cite{nauty}, a  label assigned by a function $\mathcal{C}(G)$ such that two graphs have the same label value only if they are isomorphic.

Networks often contain communities, groups of nodes more densely connected internally than with the rest of the network. Community detection algorithms~\cite{traag2019louvain} assign nodes to communities by favoring intra-community connections and limiting inter-community connections. 
Some nodes in a network may have more central positions than others. 
Centrality measures can be used to rank nodes based on their position in the network~\cite{barabasi2016network}.
The commonly used measure of betweenness centrality assigns a centrality score based on the fraction of shortest paths that pass through the node.

\subsection{Anonymity in networks}
\subsubsection{Anonymity definitions.}
We say that two nodes $v,w \in V$ are equivalent according to anonymity measure $M$ if they have the same signature: $M(v) = M(w)$.
We then say that these nodes belong to the same \emph{equivalence class}, denoted $eq_M(v)$ for a node $v$.
A node is $k$-anonymous if $|eq_M(v)| \geq k$.
We let $P_M$ denote the partition of node set $V$ into equivalence classes using measure $M$.
When $k=2$, the anonymity of a graph can be summarized using \emph{uniqueness}, defined as the fraction of unique nodes in the network with respect to measure $M$, as shown in Equation~\ref{eq:uniqueness}. 
Given this paper's focus on $k=2$, in this context network anonymity is simply $1 - U(G)$.
For later use, Equation~\ref{eq:unique-edges} also defines the set of \emph{unique edges} as the edges incident to at least one unique node.
\begin{eqnarray}
    V_{u} &=&  \{v \in V : |eq_M(v)| = 1\} \label{eq:uniquenodes}\\
    E_u &=&  \{\{v, w\} \in E : v \in V_u \vee w \in V_u\}\label{eq:unique-edges} \\
    U(G)& =& {|V_u|} / {|V|} \label{eq:uniqueness} 
\end{eqnarray}

\newpage

\subsubsection{Alteration operations.}
\label{sec:alt}
We consider three graph alteration operations: 
\begin{itemize}
    \item \textbf{Deletion}: $E' \leftarrow E \setminus \{\{v, w\}\},\ s.t. \ \{v, w\} \in E$.
    \item \textbf{Addition}: $E' \leftarrow E \cup \{\{v, w\}\},\ s.t.\ \{v, w\} \notin E$.
    \item \textbf{Rewiring}: $E' \leftarrow E \setminus \{\{v, w\}, \{v', w'\}\} \cup \{\{v, w'\}, \{v', w\}\}$ \\ \ s.t. \ $\{v, w\}, \{v', w'\} \in E$\ and\ $  \{v, w'\}, \{v', w\} \notin E$.
\end{itemize}
As preliminary experiments in Section~\ref{sub:ops} show that edge deletion is the most effective operation, we focus on edge deletion in the remainder of the text.

\subsubsection{Measures for $k$-anonymity.}\label{sub:prel-meas}
Various anonymity measures for $k$-anonymity have been introduced in the literature, differing both in the structural properties they capture and how far they reach.
In this paper, we use the measures listed in Table~\ref{tab:measures} selecting one representative measure from each category of approaches from previous work covered in Section~\ref{sec:related}.
All measures are parameterized by distance $d$, indicating up to which distance structural information is considered. 

\begin{table*}[!t]
\centering
\footnotesize
\caption{Anonymity measures (left column), signature for a given node $v$ and distance $d$ (middle column), and node set affected when altering edge $\{v,w\}$ (right column).}

\begin{tabular}{r|c|l}
\hline
Measure                                                                  & $M(v, d)$                                                                                  & $A_M(\{v, w\})$              \\ \hline
\textsc{degree}~\cite{liu2008towards,lu2012fast}                         & $degree(v)$                                                                                & $\{v, w\}$                   \\
\textsc{count}~\cite{dejong2023algorithms}                               & $(|V_{N_d(v)}|, |E_{N_d(v)} | )$                                                           & $V_{N_d(v)} \cap V_{N_d(w)}$ \\
\textsc{$d$-$k$-anonymity}~\cite{romanini2021privacy,zhou2011k} & $\mathcal{C}(N_d(v))$                                                                      & $V_{N_d(v)} \cap V_{N_d(w)}$ \\
\textsc{vrq}~\cite{hay2008resisting,thompson2009union}                   & $\{degree(u) : u \in V,$  $dist(u, v) = d\}$ & $V_{N_d(v)} \cup V_{N_d(w)}$ \\ \hline
\end{tabular}

\label{tab:measures}
\end{table*}

To determine its anonymity, each node is assigned a \emph{signature} based on the chosen measure, denoted by $M(v, d)$ in Table~\ref{tab:measures}. 
When $d\geq2$, signatures for all $d\geq1$ up to $d$ should be equal for node equivalence.
After an edge is altered, the signature of a subset of the graph's node set changes. 
Which nodes are affected depends on the \emph{reach} of the anonymity measure, i.e., how far structural information is considered to compute the node signature~\cite{dejong2024comparison}.
The rightmost column in Table~\ref{tab:measures} lists the set of affected nodes $A_M(\{v, w\})$ after deleting edge $\{v, w\}$, for each measure. 
This is used later to avoid redundant uniqueness recomputations. 

\section{The network anonymization problem}\label{sec:problem}

We now define the anonymization problem and its three variants in Definition~\ref{def:anon}.

\begin{definition}[Network anonymization problem]\label{def:anon}
    Given a network $G$, anon-ymity measure $M$ with range $d$, a value of $k$ and a set of allowed alteration operations:
    \begin{enumerate}
        \item \textbf{Full anonymization}: Make all nodes $k$-anonymous with as few alterations to the network as possible.
        \item \textbf{Partial anonymization}: Ensure that a fraction $\alpha$ (with $0.0 < \alpha < 1.0$) of the nodes is $k$-anonymous using as few alterations to the network as possible. 
        \item \textbf{Budgeted anonymization}: Perform at most $B$ (with $0 < B < |E|$) alterations to the network while maximizing the number of $k$-anonymous nodes.
    \end{enumerate}
\end{definition}
Most existing works focus on full anonymization, aiming to anonymize all nodes in the network.
However, some nodes, for example with a high degree as commonly present in real-world data, may require substantially more alterations, motivating the partial variant which generalizes the subset variant proposed in~\cite{chester2013complexity}. 
Since data utility is an important aspect of the problem~\cite{liu2008towards}, the budgeted variant limits the number of allowed alterations.

\begin{algorithm}[b!]
\begin{algorithmic}[1]
    \State \textbf{Input: graph $G$, measure $M$, value of $k$, anonymization algorithm $Anonalg$,  budget $B$, target anonymity $T$, recompute gap $R$}

    \State $G' \leftarrow G$, $G_{best} \leftarrow G$, $P_M \leftarrow compute\_P_M(G, M)$, $P_{Mbest} \leftarrow P_M$
    
    \While{$B > 0$ \textbf{and} $anonymity(G', P_M, k) < T$}\label{alg:anon:fors}
        \State $B' \leftarrow min(R, B)$        
        \State $E_A \leftarrow Anonalg(G', P_M, M, k, B')$ \label{alg:anon:alg} \Comment{Select edges to alter (Section~\ref{sub:algs})}    
        \State $G' \leftarrow update\_graph(G', E_A)$ \label{alg:anon:del}     
        \State $P_M \leftarrow update\_P_M(G', P_M, M, E_A)$ \label{alg:anon:update}
        \State $B \leftarrow B - B'$ 
        \If{$anonymity(G', P_M, k) > anonymity(G_{best}, P_{Mbest}, k)$}
            \State $G_{best} \leftarrow G'$, $P_{Mbest} \leftarrow P_{M}$
        \EndIf
    \EndWhile \label{alg:anon:fore}
    \State \textbf{Return: } $G_{best}$ \label{alg:anon:return}
    \caption{\textsc{anonymization}}
    \label{alg:anon}
\end{algorithmic}
\end{algorithm}

In this paper, we focus on $k=2$ and $d=1$, as prior work shows that increasing $k$ (up to $5$) has limited effect~\cite{dejong2023effect}, while $d>1$ is less realistic and substantially reduces anonymity~\cite{dejong2024comparison}. 
Further exploration is left for future work.

\section{Approach}\label{sec:approach}
In this section we present the general framework for anonymization and introduce the five algorithms (four new, one baseline) used in the remainder of this paper.

\subsection{Anonymization algorithm}\label{sub:anon}
The framework outlined in Algorithm~\ref{alg:anon} can be used for all variants of the anonymization problem introduced in Section~\ref{sec:problem}. 
It takes as input a budget and target anonymity, i.e., the desired number of nodes that is at least $k$-anonymous. 
For the full and partial variant, the budget equals $|E|$ and the target anonymity $T$ should be set to $|V|$ and $\alpha*|V|$ respectively, where $\alpha$ equals the fraction of nodes that should be anonymous for the partial variant of the problem.
For the budgeted variant, one should give the budget $B$ and a target anonymity of $|V|$.
The other inputs required are the original graph $G$, the measure $M$ and  value $k$, cf. Definition~\ref{def:anon}. 
The final two parameters are the chosen anonymization algorithm (see Section~\ref{sub:algs}), and recompute gap $R$, which determines after how many edge alterations $P_M$ must be recomputed, further discussed in Section~\ref{sub:lazy}.

The algorithm works as follows. 
In each iteration of the while loop in lines~\ref{alg:anon:fors} to~\ref{alg:anon:fore}, $B'$ edges are altered, as selected on line~\ref{alg:anon:alg} by the chosen anonymization algorithm. 
The network is altered (line~\ref{alg:anon:del}) and the affected part of the partition (as defined in the third column of Table~\ref{tab:measures}) is updated in line~\ref{alg:anon:update}.
The remainder of the lines are bookkeeping operations to ensure the right number of edges $B'$, equal to $R$ or the remaining budget if $B<R$, is deleted each iteration and the best result, $G_{best}$, is updated.
The \emph{anonymity()} function returns how many nodes are $k$-anonymous in the altered graph.
When the while loop terminates, either budget $B$ is depleted or target anonymity $T$ is reached.
Finally, line~\ref{alg:anon:return} returns the graph with the highest anonymity found. 

\subsection{Recompute gap}\label{sub:lazy}
Altering edges changes the network structure, which in turn influences the node signatures, and thus anonymity.
After each alteration to the graph, updating the corresponding equivalence classes and data structures is computationally expensive. 
Therefore, we incorporate a trade-off between accuracy and computational cost by updating the graph and recomputing only after $R$ alterations. 
With recompute gap $R=1$ the graph and partition are updated after each alteration, i.e., no recompute gap. For $R=B/100$ the values are updated 100 times.
Results in Appendix~\ref{app:recompute} show that using a recompute gap of 1 yields similar anonymization performance, but with substantially higher runtimes.

\subsection{Anonymization algorithms}\label{sub:algs}
The heuristic anonymization algorithms described below each assign a probability to edges to be deleted based on specific criteria. 
We focus on edge deletion as initial experiments in Section~\ref{sub:ops} showed that this yielded the best results.  
This is to be expected, as removing edges reduces the size of node neighborhoods. Smaller neighborhoods admit fewer possible network structures, making these less likely to be unique. 
The effect of edge deletion is largest on the two incident nodes, which lose one neighbor and all triangles involving that edge.  
Other affected nodes lose a single edge from their neighborhood.
Our structure based heuristic algorithms defined below, leverage these effects.
The five proposed heuristic algorithms fall into three categories:
 random edge sampling; our baseline, two structure based heuristics, and two uniqueness based heuristics.

\subsubsection{Random edge sampling.}
The baseline algorithm, \textsc{ES}~\cite{romanini2021privacy}, randomly selects which edges to delete by assigning the same probability to each edge.
\begin{align}   
    P_{\text{\textsc{es}}}(\{v, w\}) = 1 / |E|
\end{align}
\subsubsection{Structure based.}
The \textsc{degree} heuristic aims to affect more nodes by assigning higher probabilities to edges connecting two high degree nodes. These edges are likely part of many triangles.
\begin{align}  
    P_{\text{\textsc{degree}}}(\{v, w\}) = \frac{min(degree(v), degree(w))}{\sum_{\{v', w'\} \in E} min(degree(v'), degree(w'))}
\end{align}
The \textsc{aff} heuristic computes the exact number of nodes affected by deleting an edge as defined in the rightmost column of Table~\ref{tab:measures}.
\begin{align}
    P_{\text{\textsc{aff}}}(\{v, w\}) = \frac{|A_M(\{v, w\})|}{\sum_{\{v', w'\} \in E} |A_M(\{v', w'\})|}
\end{align}

\subsubsection{Uniqueness based.}
Since primarily unique nodes need to be altered in order to increase anonymity, this category of algorithms aims to have a larger impact on unique nodes.
The first uniqueness-based algorithm, \textsc{unique}, explicitly targets unique edges, i.e., edges incident to at least one unique node (see Equation~\ref{eq:unique-edges}).
From this set, the desired number of edges $B'$ is chosen at random.
If the set of unique edges is insufficiently large, the remaining edges are selected at random from the remaining edge set.  

\begin{align} 
  P_{\text{\textsc{unique}}}(\{v, w\}) =
    \begin{cases}
      1/|E_u| & \text{if $\{v, w\} \in E_u$, $|E_u| > B'$}\\
      1     & \text{if $\{v, w\} \in E_u$, $|E_u| \leq B'$}\\
      1/|E \setminus E_u| & \text{if $\{v, w\} \notin E_u$, $|E_u| < B'$}\\
      0 & \text{otherwise}
    \end{cases}  
\end{align}

The second uniqueness based algorithm, Unique Affected (\textsc{ua}), prioritizes edges whose alteration affects a larger number of unique nodes.
To ensure no edge is assigned a selection probability of zero, a factor of $\frac{1}{|E|}$ is added.
Similar to the \textsc{aff} heuristic, this heuristic is more expensive to compute when using the \textsc{count} anonymity measure considered in this paper, as the number of affected nodes need to be computed for each edge.

\begin{align}
    P_{\text{\textsc{ua}}}(\{v, w\}) = \frac{|A_{{M}}(\{v, w\}) \cap V_u| + 1/|E|}{\sum_{\{v', w'\} \in E} (|A_{M}(\{v', w'\}) \cap V_u| + 1/|E|)}
\end{align}

\subsection{Utility}\label{sub:util-metrics}
In this work, we choose to measure data utility based on the change in topological network properties and performance on common network analysis tasks: 

\begin{itemize}
    \item \textbf{Clustering coefficient.} The average clustering coefficient over all nodes. This value will likely decrease as triangles are destroyed by deleting edges.
    \item \textbf{Average distance.} The average shortest path length $dist(v, w)$ over all pairs of nodes in the same component.
    As edges are deleted, paths are destroyed, hence the average shortest path length increases. 
    However, when the network starts to split into components, this value can decrease.
    \item \textbf{Largest connected component (LCC).} The fraction of nodes in the giant component. 
    This is expected to decrease as edges are deleted. 
    \item \textbf{Centrality.} Overlap in top 100 most central nodes according to betweenness centrality before and after anonymization. If the top 100 nodes did not change the value equals 1.0, whereas it is 0.0 if there are no common nodes.
    \item \textbf{Community structure.} Normalized Mutual Information (NMI)~\cite{lancichinetti2012consensus} between the communities found by the Leiden algorithm~\cite{traag2019louvain}, before and after anonymization. To account for non-determinism of the community detection algorithm, we report on $NMI_{utility} = \max(0.0,$ $NMI_{stab} - NMI_{anon})$. 
Here $NMI_{stab}$ equals the average NMI comparing the community assignments $C$ of the original network, and $NMI_{anon}$ the average NMI between community assignments in the network before ($C$) and after ($C'$) anonymization; a value of 1.0 implies community structure is preserved, 0.0 that it is completely lost.
\end{itemize}

\section{Results}\label{sec:results}
In this section we first summarize the experimental setup and network datasets.
Then, two initial experiments compare the effectiveness of different alteration operations on graph models, and the effect of different anonymity measures on real-world network data.
Next, we compare the proposed anonymization algorithms based on their effectiveness and achieved utility on the three variants.
Finally, we compare the runtimes of the algorithms for the different variants.

\subsection{Experimental setup and data}\label{sec:data}
We implement the algorithms introduced in Section~\ref{sec:approach} in our reusable C++ ANO-NET framework.\footnote{\url{https://github.com/RacheldeJong/ANONET}}
Experiments on graph models use Python and NetworkX~\cite{networkx}, utility analysis is performed with igraph~\cite{igraph}.
Experiments are conducted on both graph models and real-world networks.
We use three graph models: Erd\H{o}s-Rényi (ER), where edges are added at random; Barab{\'a}si-Albert (BA), accounting for the preferential attachment mechanism that generates a powerlaw degree distribution, and Watts-Strogatz (WS) with rewire probability 0.05, which generates clustered networks~\cite{barabasi2016network}.
Each model uses graphs with 500 nodes and average degrees 2, 4, 16 and 64.
Properties of the real-world network datasets used can be found in Table~\ref{tab:data}.

In each experiment, edges are sequentially deleted until the desired anonymity level is reached, or the runtime exceeds 30 minutes. 
In the latter case, anonymization continues until a budget of $B=5\%$ of the edges is deleted, corresponding to the budgeted variant.
In partial anonymization, we set target anonymity $\alpha$ to $95\%$ of the nodes. 
The recompute gap is set to delete 1\% of the edges in each iteration. 
To account for nondeterminism in the algorithms and graph models, we report averages and show error bars indicating standard deviations over five runs and five generated graph instances.

For utility analysis, we generate anonymized graphs for the 19 networks that finished within the time limit by deleting the selected edges and compute the utility metrics summarized in Section~\ref{sub:util-metrics}.
For community detection, we run the Leiden algorithm~\cite{traag2019louvain} 10 times for each original and anonymized network.

Experiments are conducted on a machine with 512 GB RAM, 128 AMD EPYC 7702 cores at 2.0 GHz and 256 threads. 
Each run uses one thread, which is not shared with other processes.

\begin{table}[t!]
    \centering
    \small
    \caption{Real-world networks used, listing the number of nodes, edges, average degree, average clustering coefficient, fraction of nodes in the largest connected component (LCC) and average distance. Results include the initial uniqueness using \textsc{count}; values below 0.05 (partially anonymous networks) are shown in italics.
    The improvement ratio for the three variants is defined as the uniqueness achieved by \textsc{es} divided by that achieved by \textsc{ua}. Results not included in Section~\ref{sub:var1} are indicated by a dash.}
    \label{tab:data}
    \begin{tabular}{rrrrrrrrrrr}
    \hline
                           &                           &                                                                             &                                                                          &                                                                          &                                                                           &                            & &\multicolumn{3}{c}{\begin{tabular}[c]{@{}c@{}}Improvement \\ \textsc{UA} vs. \textsc{ES}\end{tabular}}                \\
                    & \multicolumn{1}{c}{$|V|$} & \multicolumn{1}{c}{$|E|$} & \multicolumn{1}{c}{\begin{tabular}[c]{@{}c@{}}Avg. \\ deg.\end{tabular}} & \multicolumn{1}{c}{\begin{tabular}[c]{@{}c@{}}Clust. \\ coeff\end{tabular}} & \multicolumn{1}{c}{\begin{tabular}[c]{@{}c@{}}Frac. \\ LCC\end{tabular}}  & \multicolumn{1}{c}{\begin{tabular}[c]{@{}c@{}}Avg. \\ dist.\end{tabular}} & \multicolumn{1}{c}{$U(G)$} & \multicolumn{1}{c}{\begin{tabular}[c]{@{}c@{}}1.\\ Full\end{tabular}} & \multicolumn{1}{c}{\begin{tabular}[c]{@{}c@{}}2.\\ Partial\end{tabular}} & \multicolumn{1}{c}{\begin{tabular}[c]{@{}c@{}}3.\\ Bud-\\ geted\end{tabular}} \\ \hline
Radoslaw emails~\cite{kunegis2013konect}      & 167                       & 3,250                     & 38.92                                                                    & 0.69                                                                        & 1.00                                                                                                                                            & \multicolumn{1}{r|}{1.97}                                                 & \multicolumn{1}{r|}{0.766}          & 8.0                                                                   & 2.3                                                                      & 1.9                                                                           \\
Primary school~\cite{sociopatterns}      & 242                       & 8,317                     & 68.74                                                                    & 0.53                                                                        & 1.00                                                                                                                                             & \multicolumn{1}{r|}{1.73}                                                 & \multicolumn{1}{r|}{0.975}          & 4.5                                                                   & 1.8                                                                      & 1.0                                                                           \\
Moreno innov.~\cite{kunegis2013konect}       & 241                       & 923                       & 7.66                                                                     & 0.31                                                                        & 0.49                                                                                                                                             & \multicolumn{1}{r|}{2.47}                                                 & \multicolumn{1}{r|}{0.245}          & 6.2                                                                   & 1.8                                                                      & 2.1                                                                           \\
Gene fusion~\cite{kunegis2013konect}         & 291                       & 279                       & 1.92                                                                     & 0.00                                                                        & 0.08                                                                                                                                             & \multicolumn{1}{r|}{3.90}                                                 & \multicolumn{1}{r|}{\textit{0.024}} & 4.8                                                                   & 1.0                                                                      & 5.6                                                                           \\
Copnet calls~\cite{sapiezynski2019copenhagen}        & 536                       & 621                       & 2.32                                                                     & 0.25                                                                        & 0.65                                                                                                                                            & \multicolumn{1}{r|}{7.37}                                                 & \multicolumn{1}{r|}{\textit{0.024}} & 5.5                                                                   & 1.0                                                                      & 8.6                                                                           \\
Copnet sms~\cite{sapiezynski2019copenhagen}          & 568                       & 697                       & 2.45                                                                     & 0.22                                                                        & 0.80                                                                                                                                            & \multicolumn{1}{r|}{7.32}                                                 & \multicolumn{1}{r|}{\textit{0.026}} & 2.4                                                                   & 1.0                                                                      & 3.1                                                                           \\
Copnet FB~\cite{sapiezynski2019copenhagen}           & 800                       & 6,418                     & 16.05                                                                    & 0.32                                                                        & 1.00                                                                                                                                            & \multicolumn{1}{r|}{2.98}                                                 & \multicolumn{1}{r|}{0.488}          & 7.6                                                                   & 2.1                                                                      & 1.4                                                                           \\
FB Reed98~\cite{networksrepository}           & 962                       & 18,812                    & 39.11                                                                    & 0.33                                                                        & 1.00                                                                                                                                             & \multicolumn{1}{r|}{2.46}                                                 & \multicolumn{1}{r|}{0.778}          & 5.3                                                                   & 2.3                                                                      & 1.0                                                                           \\
Arenas email~\cite{kunegis2013konect}        & 1,133                     & 5,451                     & 9.62                                                                     & 0.25                                                                        & 1.00                                                                                                                                            & \multicolumn{1}{r|}{3.61}                                                 & \multicolumn{1}{r|}{0.230}          & 8.4                                                                   & 1.7                                                                      & 2.4                                                                           \\
Euroroads~\cite{kunegis2013konect}           & 1,174                     & 1,417                     & 2.41                                                                     & 0.02                                                                        & 0.03                                                                                                                                            & \multicolumn{1}{r|}{18.37}                                                & \multicolumn{1}{r|}{\textit{0.003}} & 1.8                                                                   & 1.0                                                                      & 3.7                                                                           \\
Air traffic control~\cite{kunegis2013konect} & 1,226                     & 2,408                     & 3.93                                                                     & 0.07                                                                        & 1.00                                                                                                                                            & \multicolumn{1}{r|}{5.93}                                                 & \multicolumn{1}{r|}{\textit{0.042}} & 5.4                                                                   & 1.0                                                                      & 5.6                                                                           \\
Network science~\cite{kunegis2013konect}     & 1,461                     & 2,742                     & 3.75                                                                     & 0.88                                                                        & 0.00                                                                                                                                            & \multicolumn{1}{r|}{5.82}                                                 & \multicolumn{1}{r|}{\textit{0.039}} & 10.0                                                                  & 1.0                                                                      & 44.4                                                                          \\
FB Simmons81~\cite{networksrepository}        & 1,518                     & 32,988                    & 43.46                                                                    & 0.33                                                                        & 0.99                                                                                                                                            & \multicolumn{1}{r|}{2.57}                                                 & \multicolumn{1}{r|}{0.785}          & 2.4                                                                   & 2.2                                                                      & 1.4                                                                           \\
DNC emails~\cite{kunegis2013konect}          & 1,866                     & 4,384                     & 4.70                                                                     & 0.59                                                                        & 0.98                                                                                                                                             & \multicolumn{1}{r|}{3.37}                                                 & \multicolumn{1}{r|}{0.092}          & 15.5                                                                  & 1.5                                                                      & 1.4                                                                           \\
Moreno health~\cite{kunegis2013konect}       & 2,539                     & 10,455                    & 8.24                                                                     & 0.15                                                                        & 1.00                                                                                                                                            & \multicolumn{1}{r|}{4.56}                                                 & \multicolumn{1}{r|}{0.054}          & 6.7                                                                   & 1.0                                                                      & 3.6                                                                           \\
US power grid~\cite{kunegis2013konect}       & 3,783                     & 14,124                    & 7.47                                                                     & 0.28                                                                        & 1.00                                                                                                                                            & \multicolumn{1}{r|}{3.57}                                                 & \multicolumn{1}{r|}{\textit{0.008}} & 5.2                                                                   & 1.0                                                                      & 3.3                                                                           \\
Bitcoin alpha~\cite{networksrepository}       & 4,941                     & 6,594                     & 2.67                                                                     & 0.11                                                                        & 1.00                                                                                                                                           & \multicolumn{1}{r|}{18.99}                                                & \multicolumn{1}{r|}{0.113}          & 66.5                                                                  & 1.5                                                                      & 1.9                                                                           \\
GRQC collab.~\cite{snapnets}        & 5,241                     & 14,484                    & 5.53                                                                     & 0.69                                                                        & 0.79                                                                                                                                            & \multicolumn{1}{r|}{6.05}                                                 & \multicolumn{1}{r|}{0.054}          & 42.2                                                                  & 1.4                                                                      & 3.1                                                                           \\
Pajek Erdős~\cite{kunegis2013konect}         & 6,927                     & 11,850                    & 3.42                                                                     & 0.40                                                                        & 1.00                                                                                                                                            & \multicolumn{1}{r|}{3.78}                                                 & \multicolumn{1}{r|}{\textit{0.044}} & 55.7                                                                  & 1.0                                                                      & 4.0                                                                           \\
FB GWU54~\cite{networksrepository}            & 12,193                    & 469,528                   & 77.02                                                                    & 0.22                                                                        & 1.00                                                                                                                                            & \multicolumn{1}{r|}{2.83}                                                 & \multicolumn{1}{r|}{0.682}          & -                                                                     & -                                                                        & 1.1                                                                           \\
Enron email~\cite{snapnets}         & 36,692                    & 183,831                   & 10.02                                                                    & 0.72                                                                        & 0.92                                                                                                                                            & \multicolumn{1}{r|}{4.03}                                                 & \multicolumn{1}{r|}{0.071}          & -                                                                     & -                                                                        & 2.0                                                                           \\
FB wall 2009~\cite{kunegis2013konect}        & 45,813                    & 183,412                   & 8.01                                                                     & 0.15                                                                        & 0.96                                                                                                                                            & \multicolumn{1}{r|}{5.60}                                                 & \multicolumn{1}{r|}{\textit{0.033}} & -                                                                     & -                                                                        & 4.6                                                                           \\
Brightkite~\cite{networksrepository}          & 58,228                    & 214,078                   & 7.35                                                                     & 0.27                                                                        & 0.97                                                                                                                                           & \multicolumn{1}{r|}{4.92}                                                 & \multicolumn{1}{r|}{\textit{0.048}} & -                                                                     & -                                                                        & 2.6        \\
Twitter~\cite{kunegis2013konect} & 465,017 & 833,540 & 3.59 & 0.06 & 1.00 & \multicolumn{1}{r|}{4.59} & \multicolumn{1}{r|}{\textit{0.004}} & - & - & 1.4
\\ \hline 
\end{tabular}
\end{table}

\subsection{Alteration operations and anonymity}\label{sub:ops}
\begin{figure}[b!]
	\centering
    \vspace{-1em}
    \includegraphics[width=\textwidth]{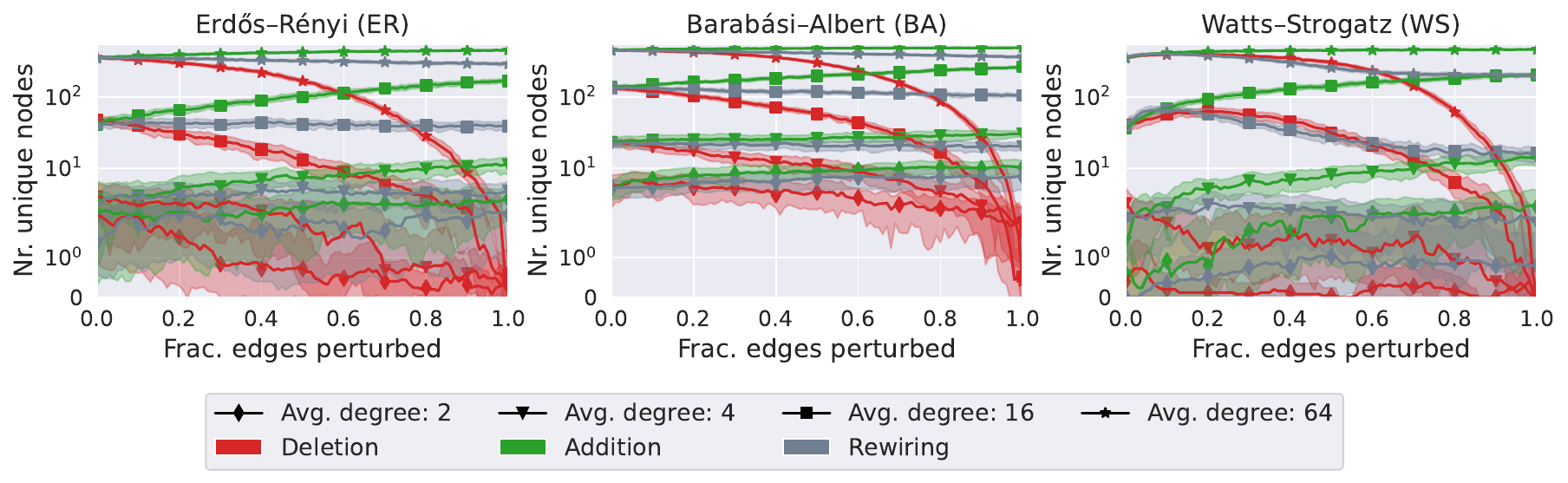}
    \vspace{-1em}
	\caption{Random edge deletion (red), addition (green),  and rewiring (grey) applied to, using the \textsc{count} measure, anonymize ER (left), BA (middle) and WS (right) graphs with $|V| = 500$ and average degree $\in \{2, 4, 16, 64\}$.}
	\label{fig:resoperations}
\end{figure}

Figure~\ref{fig:resoperations} shows results on how repeatedly applying each operation discussed in Section~\ref{sec:alt} affects the anonymity of graph models.
First note that the number of unique nodes when no edges are altered varies with the average degree. 
Graphs with higher average degree have a higher initial uniqueness, which corresponds to findings in earlier work~\cite{dejong2023effect,romanini2021privacy}.
Clearly, random deletion is more effective than random addition or rewiring, confirming what we theorized in Section~\ref{sub:algs}.
Interestingly, for the graph models, random addition and rewiring can even result in a higher uniqueness, as shown for the BA model, and the ER and WS models when the average degree is larger than 16.
However, for the WS model, with average degree 16, rewiring performs slightly better than deletion when altering a fraction of 0.2 to 0.5 of the edges.
This is likely due to the many nearly identical near-cliques in the generated networks.

While these preliminary results can not rule out that edge addition and rewiring can be more effective and perhaps preferred in specific cases, for example when targeting specific edges, we choose to focus on edge deletion in the remainder of the paper, as it unequivocally performed best on average.

\subsection{Anonymization and measures}\label{sub:anonmeas}
\begin{figure*}[t!]
	\centering
    \includegraphics[width=0.9
\textwidth]{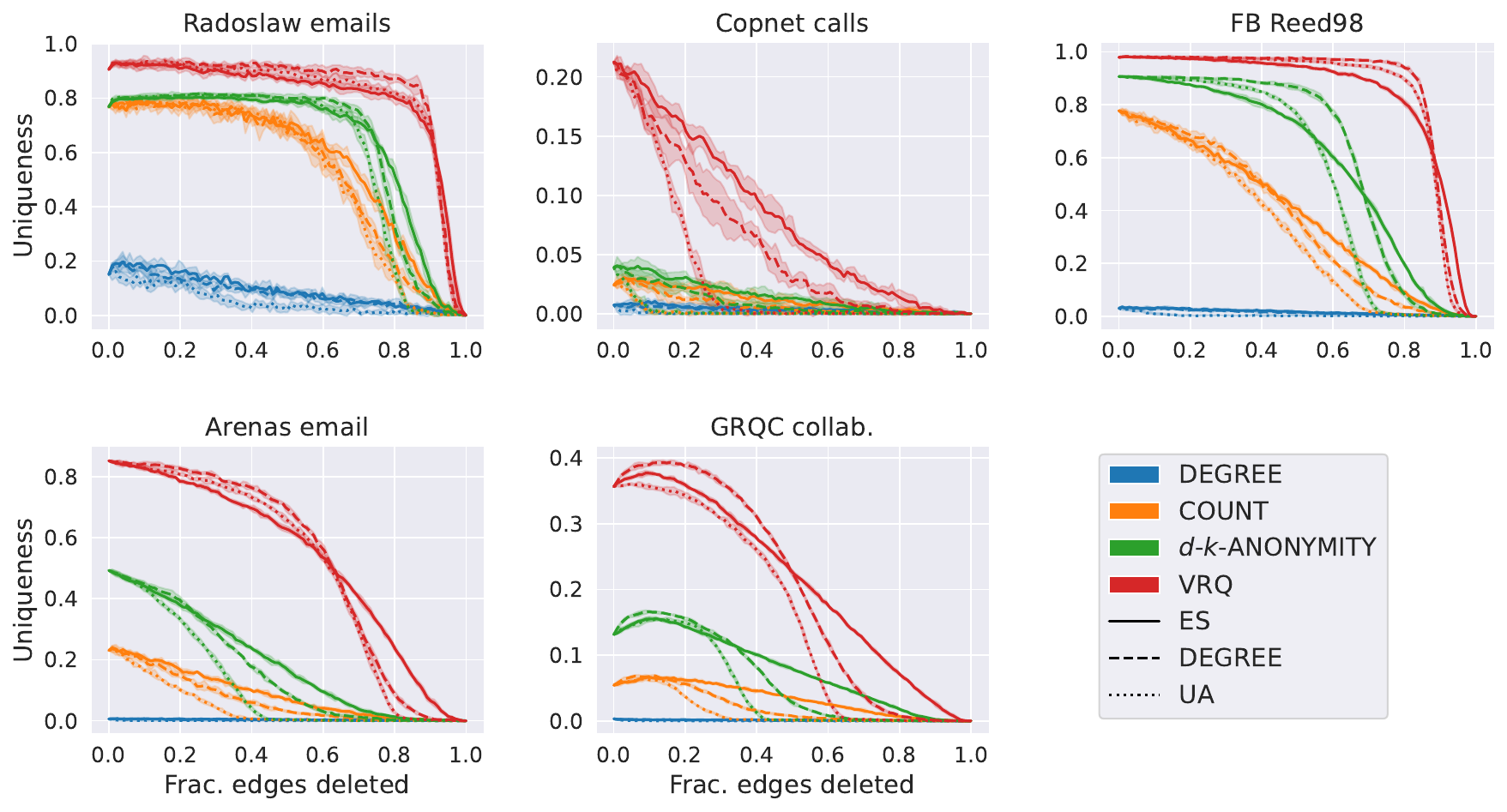}	
	\caption{Uniqueness for four different anonymity measures (color) and three anonymization algorithms (linestyle) on five real-world networks. For each network we show the fraction of deleted edges (horizontal axis) and the attained uniqueness (vertical axis).}
	\label{fig:resmeasures}
\end{figure*}

Figure~\ref{fig:resmeasures} shows how uniqueness changes when deleting edges using different measures for anonymity and each algorithm in Section~\ref{sub:algs}.
We depict a selection of five networks representative of the overall observed behavior for the networks listed in Table~\ref{tab:data}. 
This is, from top to bottom, left to right in Figure~\ref{fig:resmeasures}, 1) a plateau with a steep decrease, 2) linear decrease, 3) plateau with linear decrease, 4) sigmoid-like decrease and 5) initial increase before decrease. 
For completeness, Appendix~\ref{app:anon} contains results for the remaining networks.

Overall, we see that the measure for anonymity used influences both the initial uniqueness and how fast the uniqueness decreases.
The simplest measure, \textsc{degree}, has the lowest initial uniqueness and is the easiest to anonymize for.
The measure with the furthest reach, \textsc{vrq}, has the highest initial uniqueness and is the most difficult to anonymize for, corresponding with findings in~\cite{dejong2024comparison}.

Counterintuitively, in the ``GRQC collab.'' network (bottom right), the \textsc{es} algorithm shows an increase in uniqueness before decreasing.
This is likely due to the collaboration network's many cliques.
When randomly deleting edges, these cliques are destroyed, which may make certain nodes unique.
If edges are targeted more specifically, for example, using the \textsc{ua} heuristic, this initial increase does not occur.

In the remainder of this paper, we focus on the \textsc{count} measure, as this models a realistic attacker scenario, has a substantial initial uniqueness and is easier to anonymize for than, for example, the \textsc{$d$-$k$-anonymity} measure. 

\subsection{Full, partial and budgeted network anonymization}\label{sub:var1}
In this section, we investigate how effective the proposed algorithms from Section~\ref{sub:algs} are at solving the three variants of the network anonymization problem.

\begin{figure*}[t!]
	\centering
    \includegraphics[width=\textwidth]{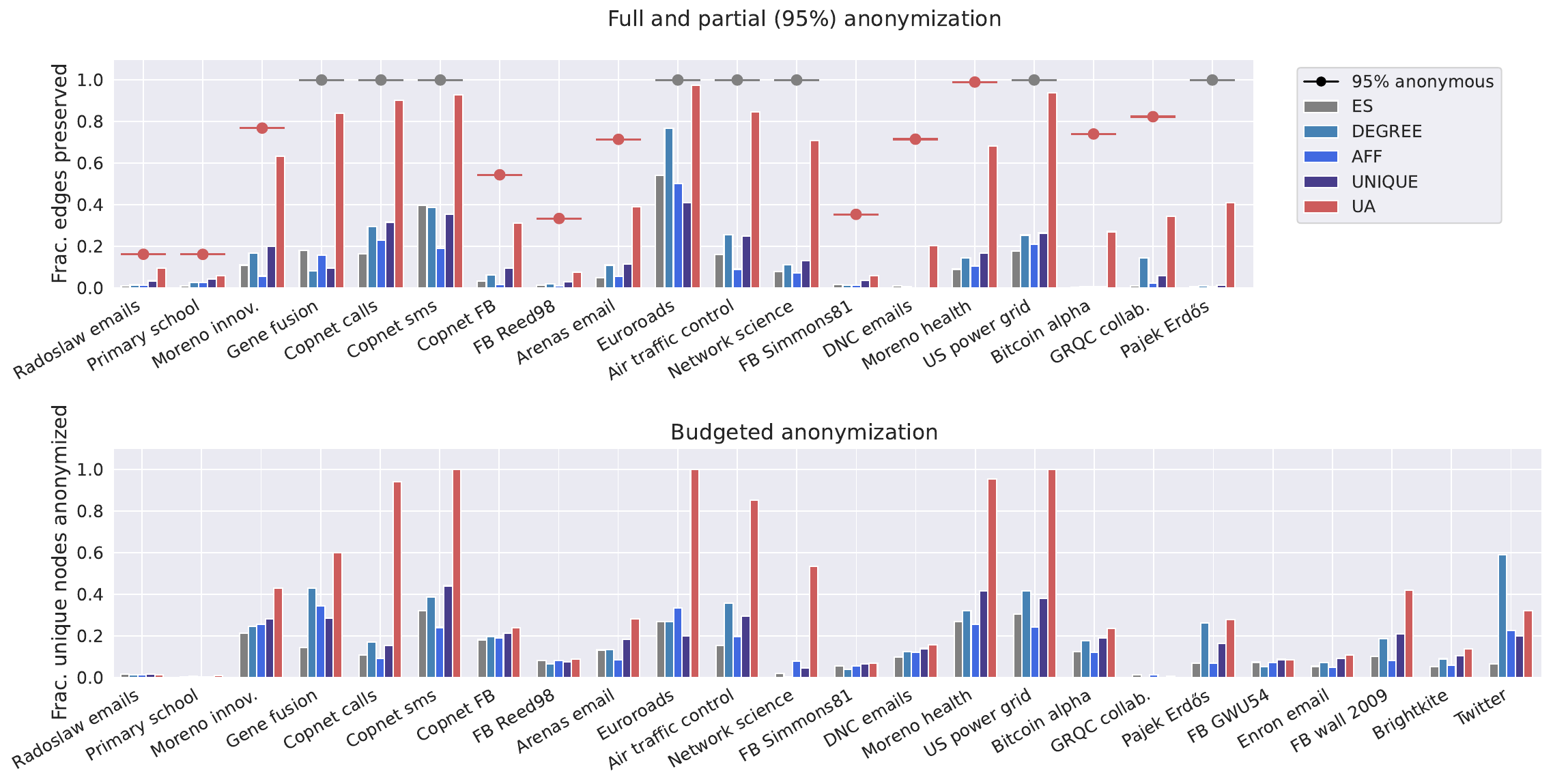}
	\caption{Top: Results for full and partial anonymization. Bars indicate the fraction of edges preserved in full anonymization, horizontal lines the fraction of edges preserved for partial anonymization. 
    Bottom: Bars indicate the fraction of unique nodes anonymized within the budget of $B=0.05|E|$. Higher values correspond to better performance.
    }
	\label{fig:resall}
\end{figure*}

The top of Figure~\ref{fig:resall} shows the fraction of edges preserved $\frac{|E'|}{|E|}$ after deleting edges until the desired level of anonymity is reached.
For the partial variant, eight of the networks, including ``US Power grid'' and ``Euroroads'' retain 100\% of their edges as initially less than 5\% of their nodes is unique.

For some networks, such as ``Radoslaw emails'' and ``FB Simmons81'' only a small fraction of edges can be preserved.
In networks where more edges can be preserved, the uniqueness based algorithm \textsc{ua} outperforms the other algorithms.
Dividing the performance of \textsc{ua} by that of \textsc{es}, we obtain the \emph{improvement ratio} reported in the three rightmost columns of Table~\ref{tab:data}.
On average, for the full and partial variants, 13.9 and 1.8 times more edges can be preserved, suggesting that even simple heuristics provide substantial improvement. 
The bottom of Figure~\ref{fig:resall} shows results for the budgeted variant.
To account for differences in initial uniqueness, the vertical axis shows the fraction of unique nodes that is anonymized, $1 - \frac{U(G')}{U(G)}$, when deleting at most 5\% of the edges.
A higher value indicates greater effectiveness within the given budget.

Overall, the \textsc{ua} algorithm performs well in anonymizing networks, particularly those with lower initial uniqueness, average degree, and higher average distance, as further supported by results in Appendix~\ref{app:prop}.
Some of the networks, such as ``Radoslaw emails'' and ``GRQC collab.'', require more than 5\% of edge deletion to achieve substantial improvements.
On average, in the budgeted variant, \textsc{ua} anonymizes 4.8 times more nodes than  \textsc{es}.

\subsection{Utility}\label{sub:util}
\begin{figure}[t!]
	\centering
    \includegraphics[width=0.95\textwidth]{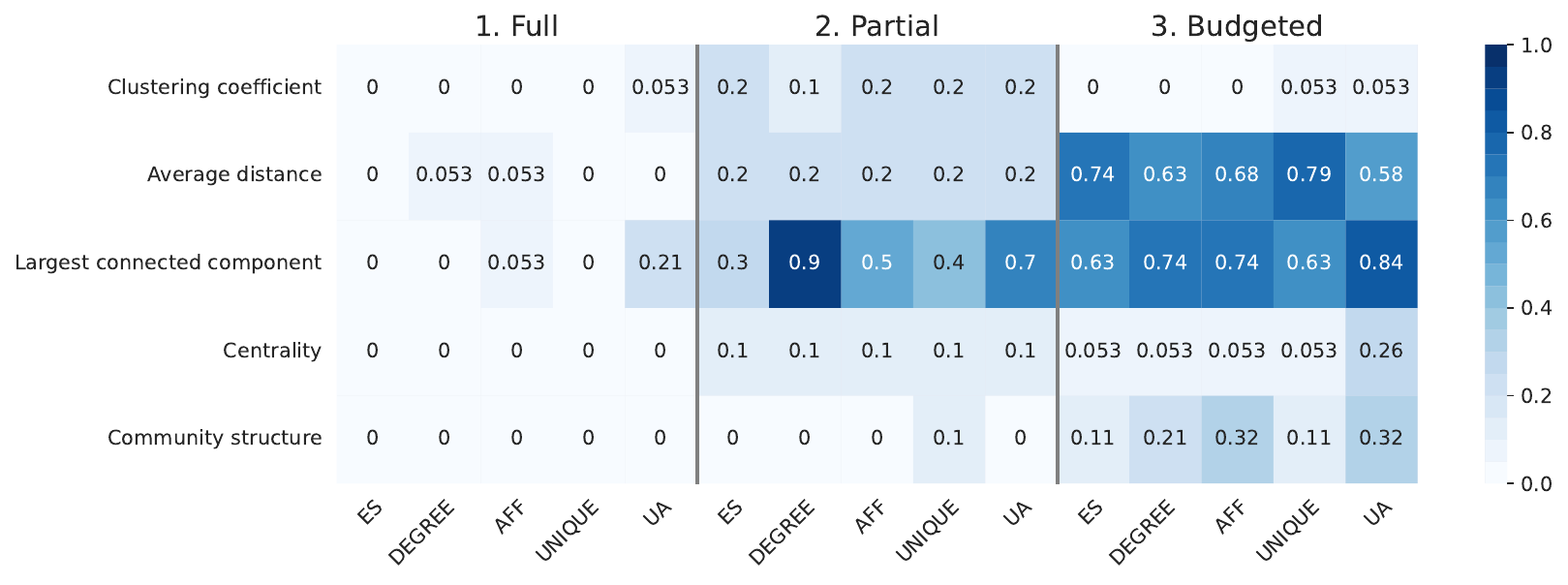}
	\caption{For each variant of the anonymization problem (columns) and each algorithm (horizontal axis), cells indicate the fraction of networks for which utility (vertical axis) is preserved, i.e., the change $\pm$ one standard deviation is less than 5\% of the original. 
    }
	\label{fig:utiltable}
\end{figure}

The utility of the anonymized networks in terms of how well various network properties and performance on network analysis tasks are preserved across the different variants of the anonymization problem is summarized in 
Figure~\ref{fig:utiltable}. 
A property is considered preserved if the value after anonymization, including one standard deviation, differs by less than 5\% from the original value.
For partial anonymization, we exclude the eight partially anonymous networks (initial uniqueness below 5\%), leaving 11 networks.  
For completeness, Appendix~\ref{app:deleting} presents utility results for the networks, showing how utility evolves during the anonymization process.

Comparing the three variants, we observe that for full and partial anonymization most properties are generally not preserved.
Exceptions are the clustering coefficient and average distance which are preserved for one network, and the largest component size, which the \textsc{ua} algorithm preserves in four networks for the full variant --- likely because \textsc{ua} requires fewer edge deletions and targets different edges.
In the budgeted variant properties are preserved more often, and differences between algorithms are smaller.
The \textsc{ua} algorithm best preserves the majority of properties, except for average shortest path length.

It is also important to note that even random edge deletion (\textsc{es}) does not always preserve network properties. 
From the considered properties, average shortest path length and largest component size are preserved most consistently. 
The latter indicates that the networks often stay connected after anonymization.
Community structure and the most central nodes are preserved for a substantial fraction of the networks.
However, the average clustering coefficient seems challenging to preserve.
These findings suggest that the budgeted variant provides a promising setting for further investigation and comparison of anonymization algorithms when data utility is of importance. 

\subsection{Runtime}\label{sub:runtime}
\begin{figure}[t!]
	\centering
    \includegraphics[width=\textwidth]{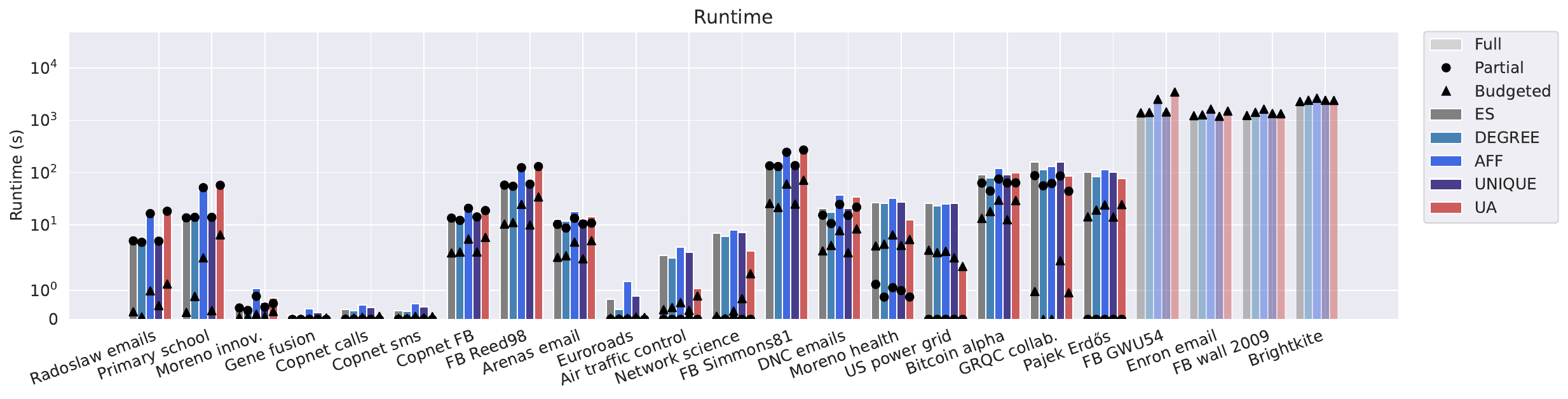}
	\caption{Runtimes for the five algorithms (color) with recompute gap $|E|/100$. Runtimes are shown for the three variants, full (bar), partial (circle) and budgeted (triangle). For the four largest networks, only runtimes of the budgeted variant are included.}
	\label{fig:runtime}
\end{figure}

Figure~\ref{fig:runtime} shows the observed runtime of each algorithm, on each of the three variants, for each network.
The runtime consists of computing probabilities to choose edges, for which the computational complexity varies for the anonymization algorithms introduced in Section~\ref{sub:algs}, deleting the chosen edges from the network, which can be done in linear time, and accordingly determining the new anonymity of nodes by updating the equivalence partition (lines~\ref{alg:anon:del} and \ref{alg:anon:update} in Algorithm~\ref{alg:anon}).
For this last operation, which is a computationally expensive part of the anonymization process, the runtime is affected by the number of affected nodes and the density of the network.

Comparing the different problem variants, the lowest runtimes are usually observed for the the budgeted variant, except for networks that do not require edge deletions for the partial variant if their initial uniqueness is less than 5\%.
Runtimes for the partial and full variant are often very similar.
This small difference is likely due to the fact that later iterations require less runtime, which is further supported by results on the runtime per iteration in Appendix~\ref{app:runtimepart}.

Overall the runtime increases for networks with more nodes.
At the same time, the first four networks show that a high uniqueness and density increase runtimes.
The latter makes both measuring anonymity and computing the number of affected nodes more computationally expensive.
For the four largest networks, which were included to assess the scalability of the algorithms, we observed runtimes of over 30 minutes for the budgeted variant for all algorithms, including \textsc{es} which spends a minimal amount of time on selecting edges.

Moving to the different algorithms, we first note that for many of the networks, there is no large difference in the runtimes required by the algorithms, implying that even on large networks the runtime can be high due to the bookkeeping operations.
As shown in Appendix~\ref{app:runtimepart}, the time required for bookkeeping operations, including updating the equivalence classes, is usually much larger than the time used for the anonymization algorithms.
Exceptions are networks such as ``FB Simmons'' for which the \textsc{unique} and \textsc{ua} spend the largest amount of time on the anonymization algorithm itself.
However, for some networks, such as ``Air traffic control'' and ``Network science'',  the \textsc{ua} algorithm is slightly faster than the other algorithms, including \textsc{es}.
This can be explained by the fact that \textsc{ua} is more effective in anonymization, which results in fewer required iterations and a lower runtime. 

\section{Conclusion}\label{sec:conc}
In this paper, we introduced the anonymization problem in social networks and its three variants.
We introduced ANO-NET, a generic computational algorithmic framework that allows one to anonymize networks under these variants, supporting a range of anonymity measures, the proposed anonymization algorithms and options to explore relevant parameters.
Experiments on graph models showed that, among the considered edge alteration operations, random edge deletion is most effective.
Moreover, we found that stricter anonymity measures come with higher  initial uniqueness and make the anonymization process more challenging.
We proposed and evaluated five measure-agnostic anonymization algorithms. 
Their effectiveness varied across networks; those with lower initial uniqueness and higher average path length were easier to anonymize.
The \textsc{ua} algorithm, which targets edges affecting the largest number of unique nodes, consistently showed to be most effective.
Compared to the baseline, it anonymized on average 4.8 times more nodes in the budgeted variant and preserved 13.9 and 1.8 times more edges for full and partial anonymization, respectively.
Additionally \textsc{ua} better preserved several utility metrics in the budgeted variant, realizing the best overall trade-off between anonymity and data utility. 

This work lays the foundation for further studying the anonymization problem and its variants and for extending anonymity measures and algorithms to account for additional network properties such as node labels, layered edges or temporal aspects.
Finally, dynamically monitoring utility during the anonymization process to guide edge alterations represents another promising avenue for future research.

\bibliographystyle{splncs04}
\bibliography{bibliography}%

\clearpage
\appendix
\section{Recompute gap}~\label{app:recompute}
\begin{figure*}[h]
	\centering
    \includegraphics[width=0.9\textwidth]{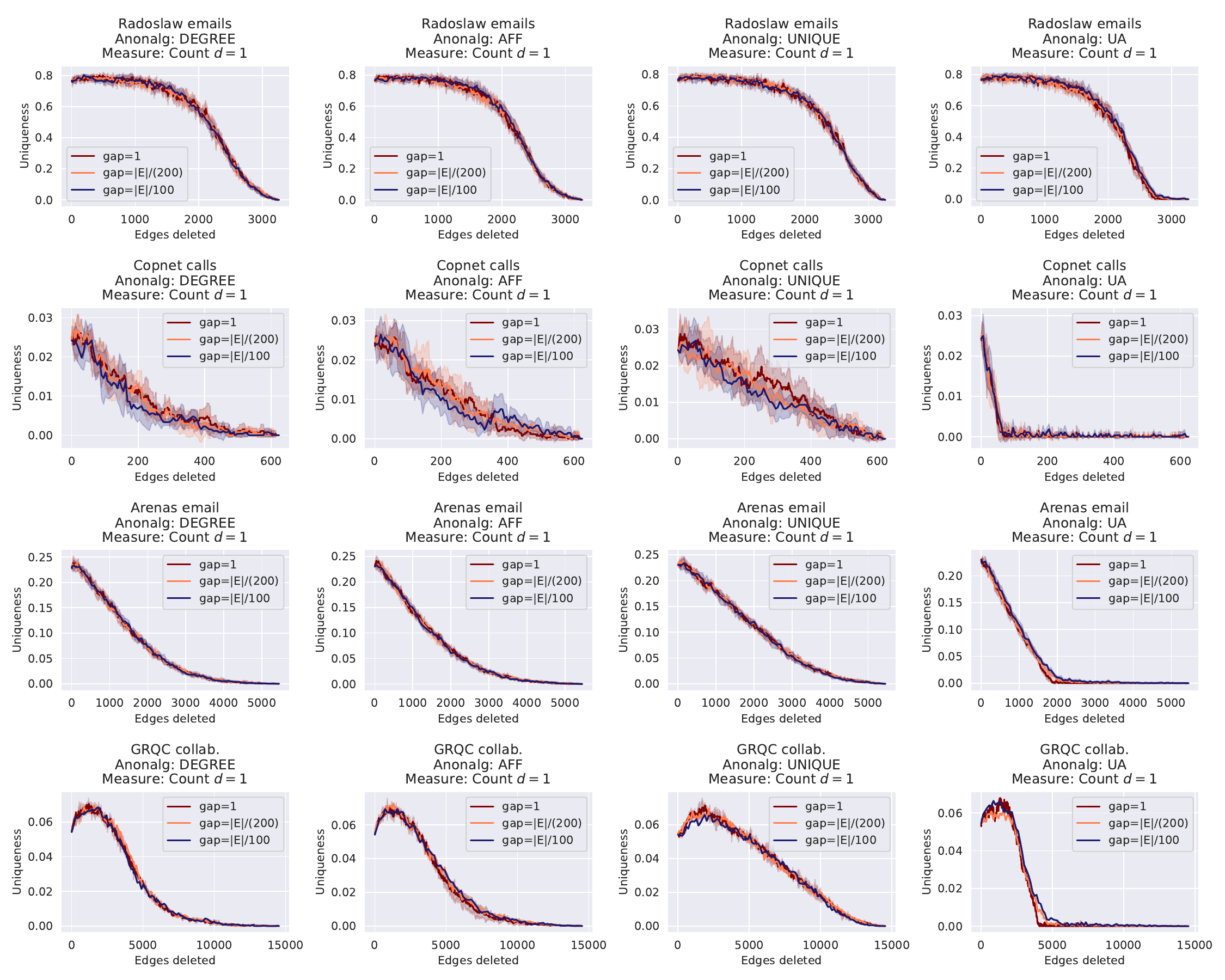}
	\caption{Anonymization process for a selection of networks, anonymization algorithms and different recompute gaps (color).}
	\label{fig:gapres}
\end{figure*}

\begin{figure*}[t]
	\centering
    \includegraphics[width=\textwidth]{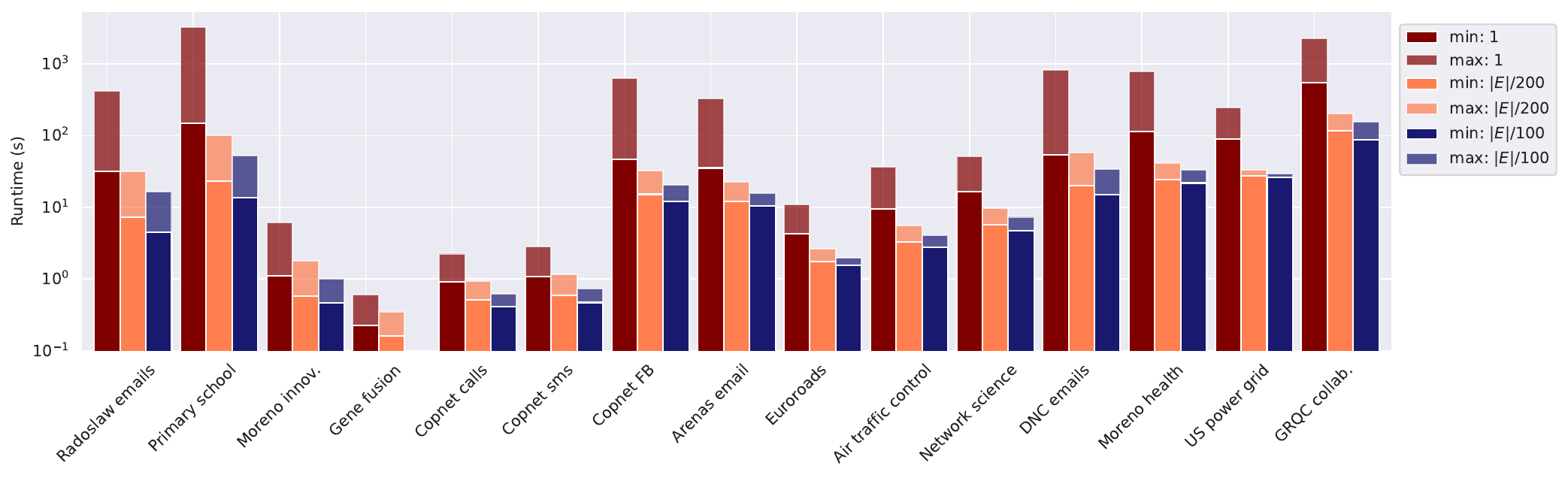}
	\caption{Runtimes for anonymization using different recompute gap values (1, $|E|$/200 and $|E|$/100). Opaque values denote the largest average runtime of the anonymization algorithms, while the solid values denote the lowest value found. }
	\label{fig:gaptime}
\end{figure*}

This appendix accompanies Section~\ref{sub:lazy} of the full paper.
Here, we compare the performance and runtimes achieved by the anonymization algorithms when using different recompute gaps.
Figure~\ref{fig:gapres} shows the results for different algorithms on a subset of the network datasets when using different recompute gaps.
The recompute gap value of 1 implies that values are recomputed after each edge deletion, $|E|/100$ equals the default setting.
The figure overall shows that using a smaller recompute gap similar results are obtained.
Turning to the runtimes in Figure~\ref{fig:gaptime} we see that using a smaller recompute gap does result in larger runtimes.

\newpage
\section{Anonymization process}~\label{app:anon}
\begin{figure*}[h!]
	\centering
    \includegraphics[width=0.925\textwidth]{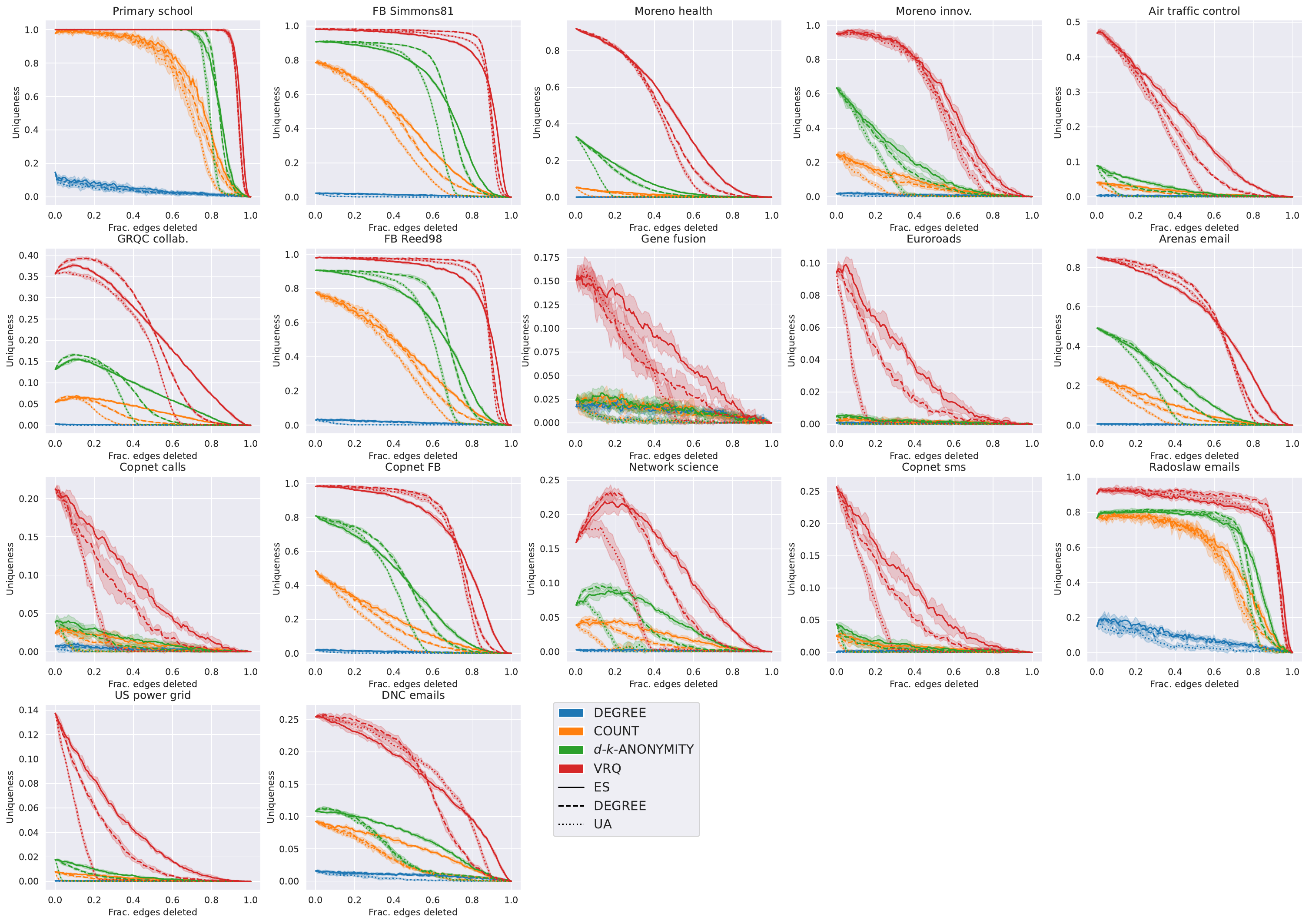}
	\caption{Anonymization process for all networks, anonymization algorithms (line style) and measures (color).}
	\label{fig:measapp}
\end{figure*}

This appendix contains supplementary figures accompanying Section~\ref{sub:anonmeas}.
Results shows the full anonymization process, i.e., up to deleting all edges.
Figure~\ref{fig:measapp} shows results for all anonymity measures for all networks not shown in Figure~\ref{fig:resmeasures} of the full paper.

\section{Properties and anonymization}\label{app:prop}
\begin{figure}[h]
	\centering
    \includegraphics[width=\textwidth]{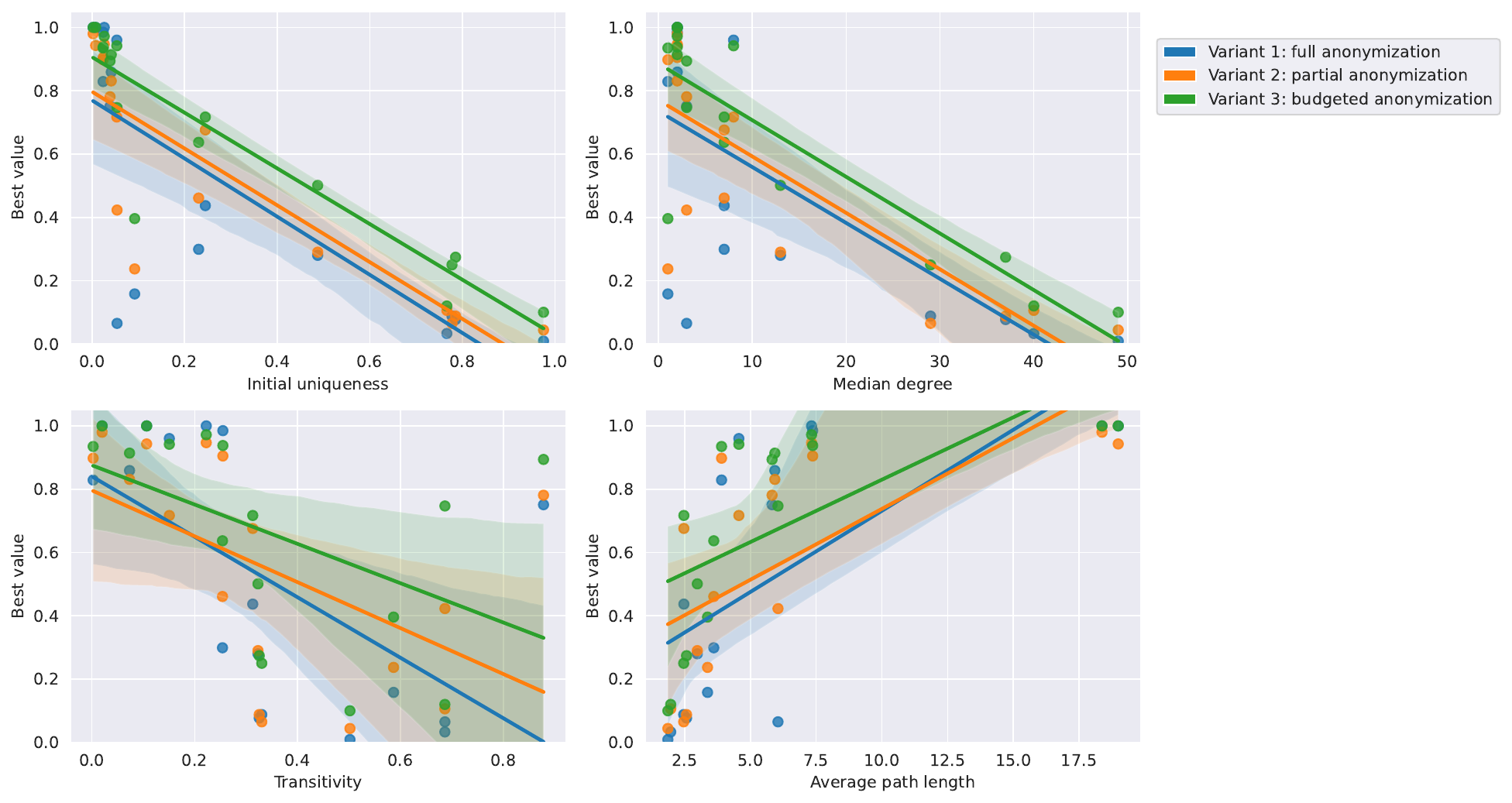}
	\caption{Network properties and anonymization. Each plot shows the the best value obtained (vertical axis) for each anonymization variant (color) on a specific network and the value of a network property (horizontal axis).
    Combinations plotted are those with a correlation deemed significant as denoted in Table~\ref{tab:propa}.}
	\label{fig:prop}
\end{figure}

This appendix accompanies Section~\ref{sub:var1} of the full paper and contains results on correlations between network properties and the effectiveness of the anonymization algorithms on the three variants of the anonymization problem.
Table~\ref{tab:propa} denotes the Pearson correlation and p-value.
For combinations deemed significant, Figure~\ref{fig:prop} shows the values for all networks.

\begin{table}[h]
    \begin{tabular}{@{}rrrrrrr@{}}
    \cmidrule(l){2-7}
                        & \multicolumn{2}{c}{1. Full}                                                                 & \multicolumn{2}{c}{2. Partial}                                                              & \multicolumn{2}{c}{3. Budgeted}                                                             \\ \cmidrule(l){2-7} 
    Network property    & \begin{tabular}[c]{@{}r@{}}Pearson\\ correlation\end{tabular} & \multicolumn{1}{l}{p-value} & \begin{tabular}[c]{@{}r@{}}Pearson\\ correlation\end{tabular} & \multicolumn{1}{l}{p-value} & \begin{tabular}[c]{@{}r@{}}Pearson\\ correlation\end{tabular} & \multicolumn{1}{l}{p-value} \\ \midrule

        Unique start        & \textbf{-0.86} & \textbf{0.00} & \textbf{-0.92} & \textbf{0.00} &\textbf{ -0.76} & \textbf{0.00} \\
        $|V|$               & 0.13  & 0.62 & 0.29  & 0.25 & 0.06  & 0.81 \\
        Average degree      & -0.19 & 0.45 & -0.13 & 0.61 & -0.15 & 0.56 \\
        Median degree       &\textbf{ -0.80} &\textbf{ 0.00 }& \textbf{-0.87} & \textbf{0.00} & \textbf{-0.68} & \textbf{0.00} \\
        Transitivity        & -0.52 & 0.03 & -0.48 & 0.05 & -0.59 & 0.01 \\
        Assortativity       & 0.15  & 0.56 & 0.29  & 0.25 & 0.10  & 0.71 \\
        diameter            & \textbf{0.64}  & \textbf{0.01}  & \textbf{0.60}  & \textbf{0.01} & \textbf{0.62 } & \textbf{ 0.01} \\
        Average path length & \textbf{0.65} & \textbf{0.00} & \textbf{ 0.62}  & \textbf{0.01} & \textbf{ 0.64}  & \textbf{0.01} \\ \hline
    \end{tabular}
    \caption{
    Graph properties and correlation with anonymization results. 
    Given various graph measures, denoted in the leftmost column, the Pearson correlation is shown with the best outcomes of the three anonymization variants (second column to last column). For each outcome both the Pearson correlation and $p$-value are given. Values with $p < 0.05$ and Pearson correlation larger than 0.4, or smaller than -0.4, are shown in bold.}
    \label{tab:propa}
\end{table}
\clearpage

\section{Deleting edges}\label{app:deleting}
This appendix accompanies Section~\ref{sub:util} of the full paper.
Each figure shows how a network property or performance on a downstream task changes when deleting more edges, using the five anonymization algorithms.

\begin{figure}[htbp]
\begin{center}
    {\includegraphics[width=\textwidth]{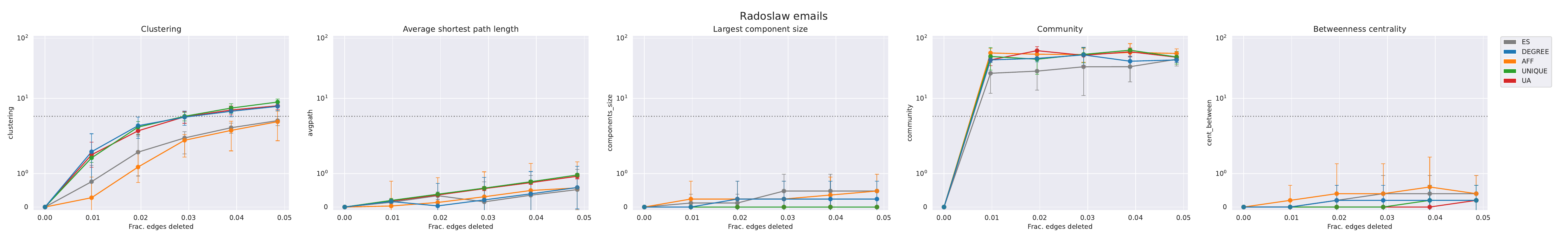}}
    {\includegraphics[width=\textwidth]{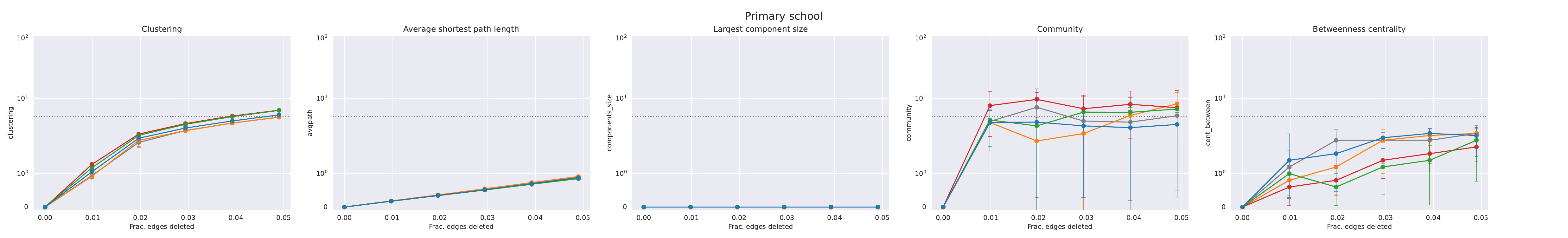}}
    {\includegraphics[width=\textwidth]{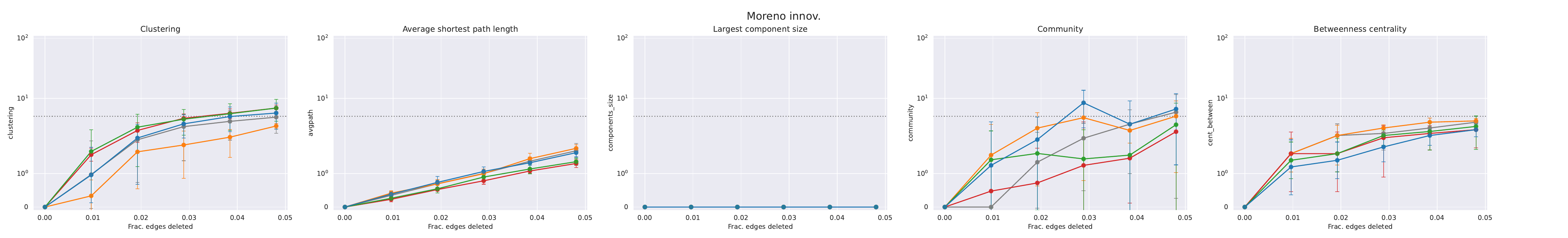}}
    {\includegraphics[width=\textwidth]{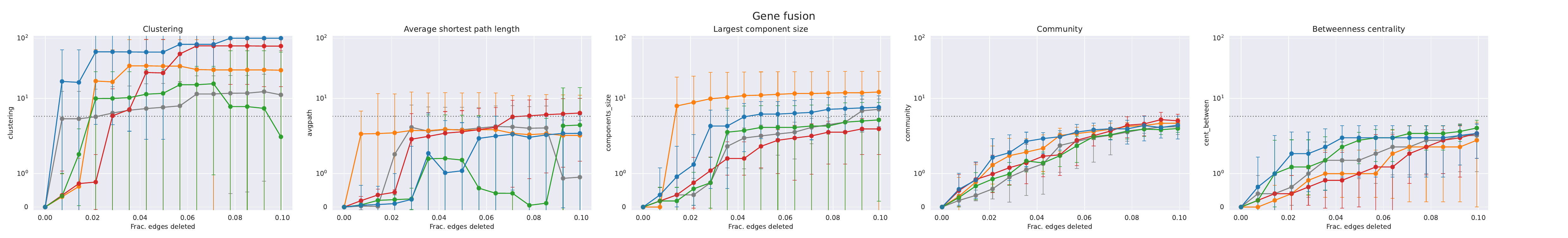}}
    {\includegraphics[width=\textwidth]{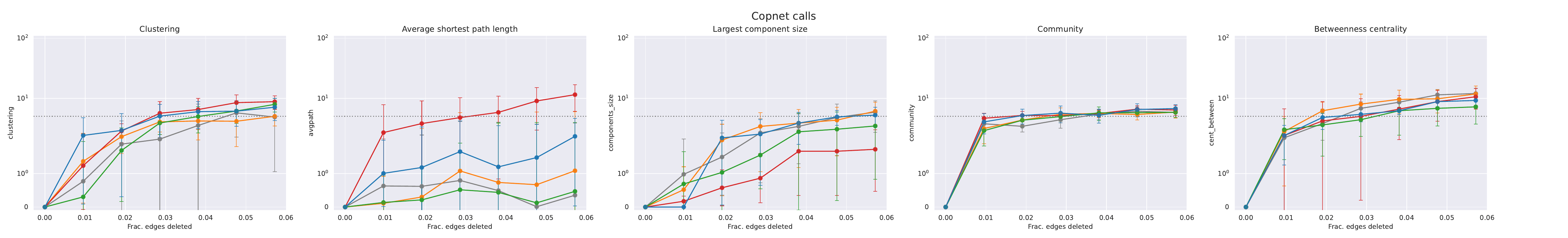}}
    {\includegraphics[width=\textwidth]{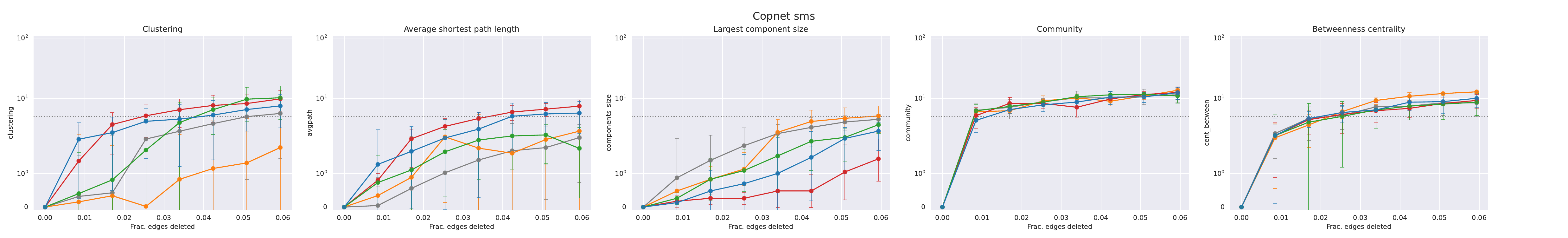}}
    {\includegraphics[width=\textwidth]{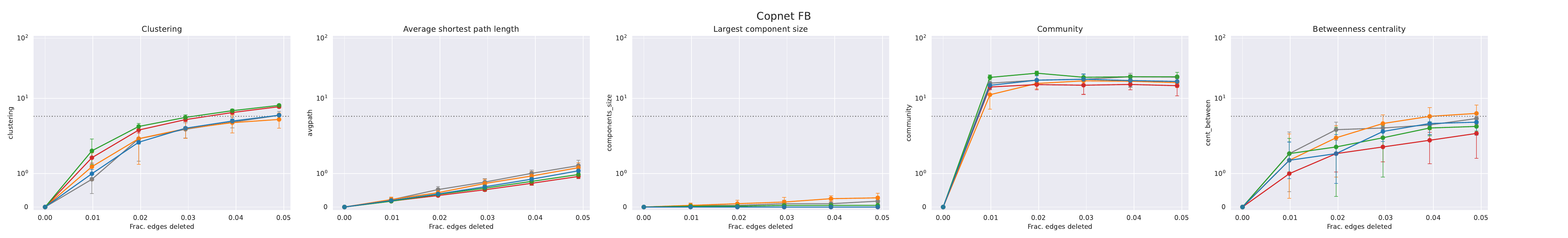}}   
\end{center}
\end{figure}

\begin{figure}[htbp]
\begin{center}    
    {\includegraphics[width=\textwidth]{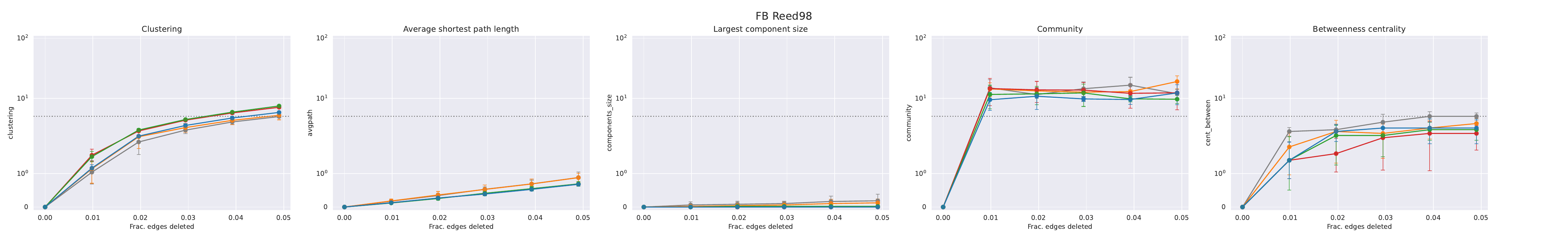}}
    {\includegraphics[width=\textwidth]{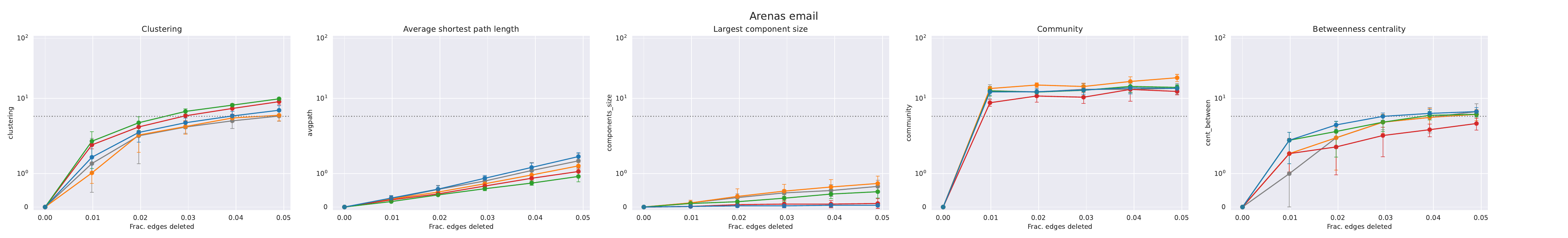}}
    {\includegraphics[width=\textwidth]{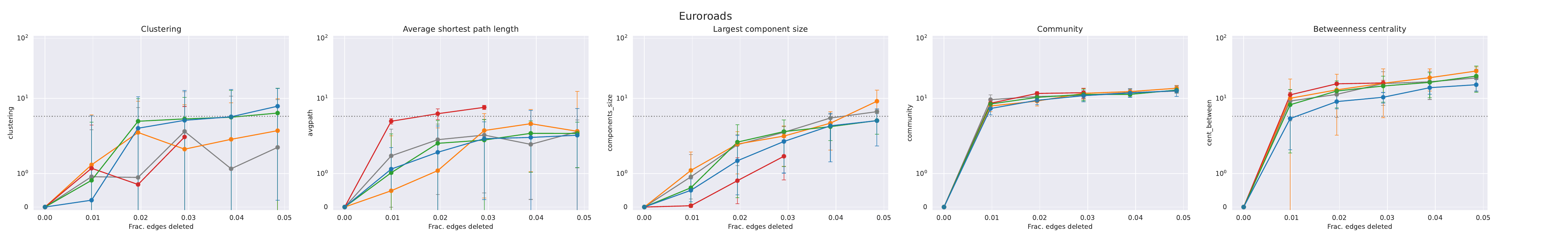}}
    {\includegraphics[width=\textwidth]{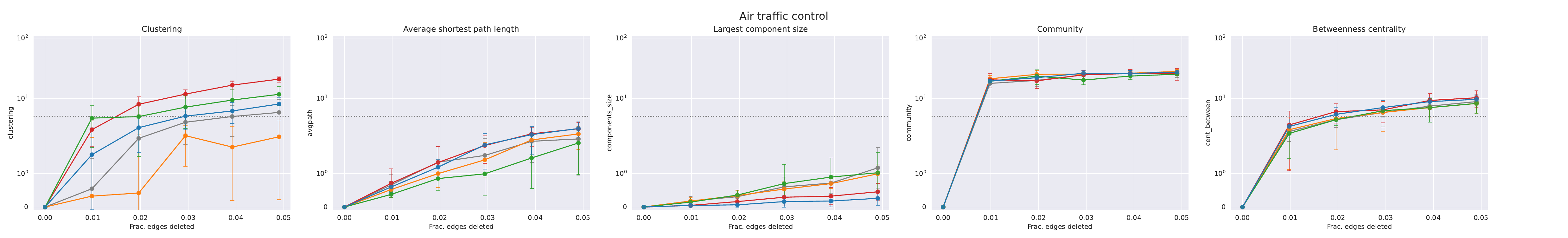}}
    {\includegraphics[width=\textwidth]{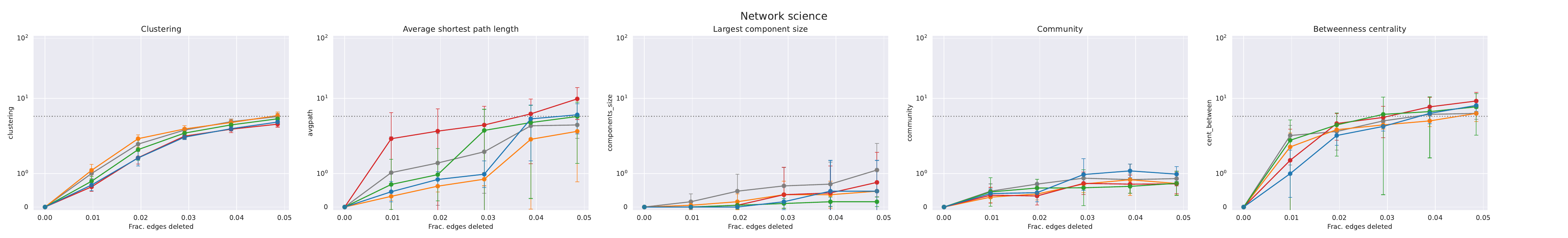}}
    {\includegraphics[width=\textwidth]{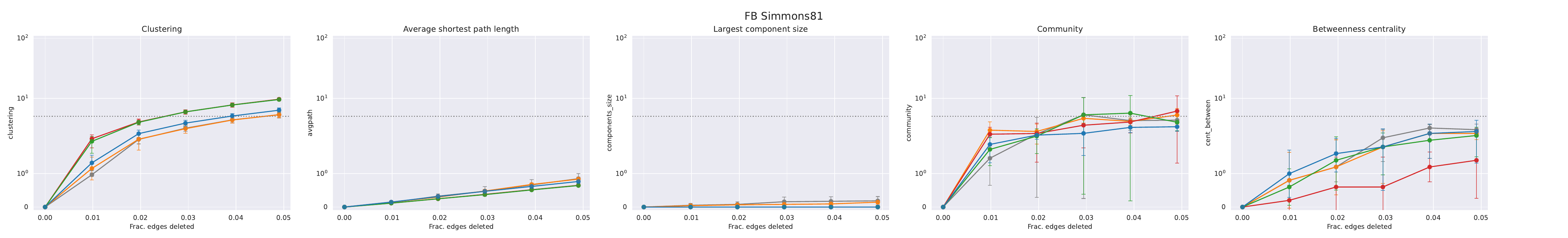}}
    {\includegraphics[width=\textwidth]{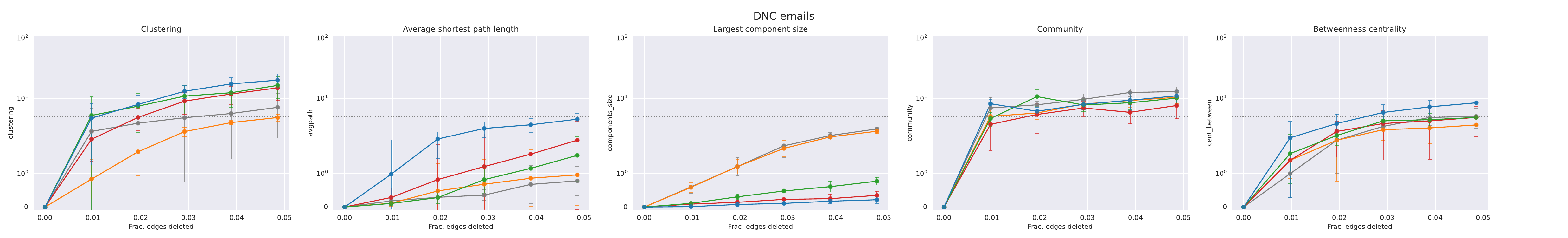}}
    {\includegraphics[width=\textwidth]{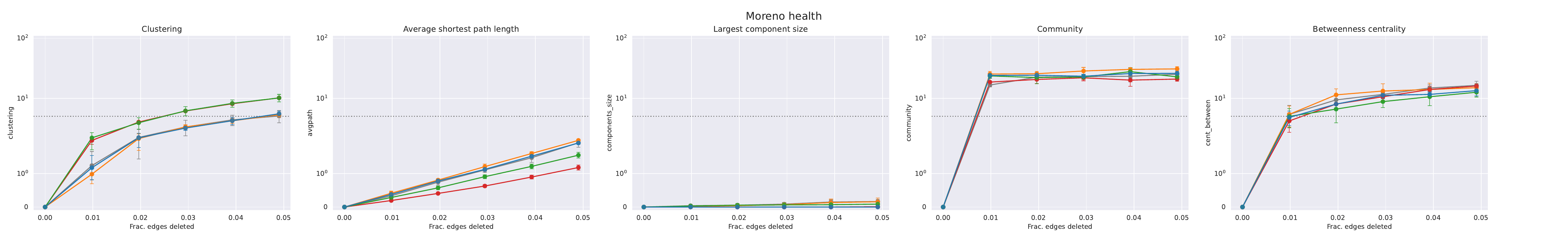}}
    {\includegraphics[width=\textwidth]{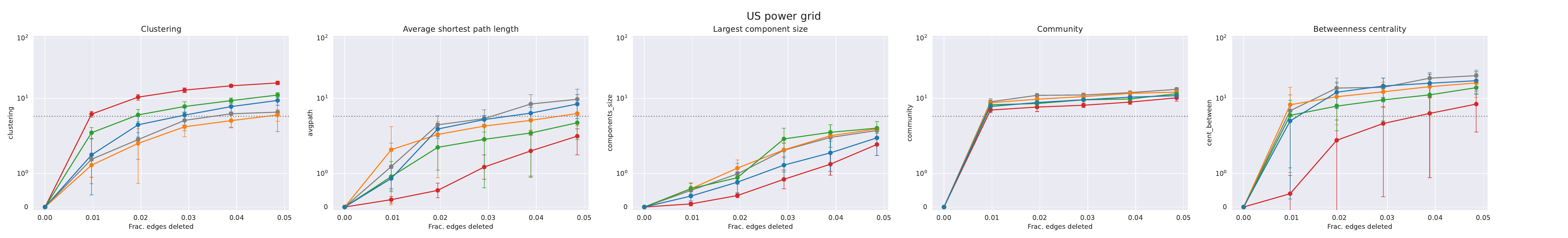}}

\end{center}
\end{figure}

\begin{figure}[htbp]
\begin{center}  
    {\includegraphics[width=\textwidth]{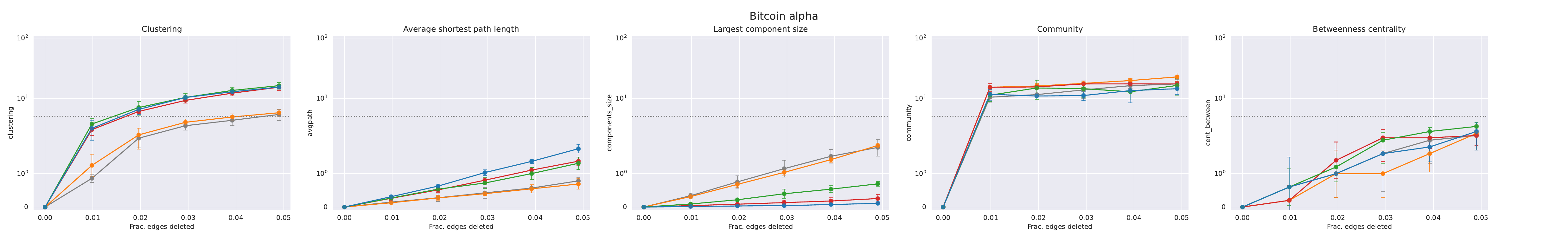}}
    {\includegraphics[width=\textwidth]{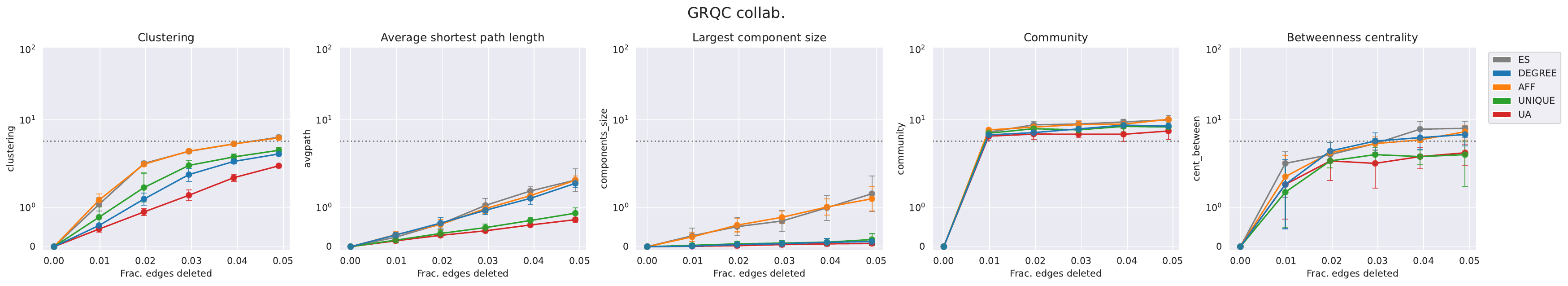}}
    {\includegraphics[width=\textwidth]{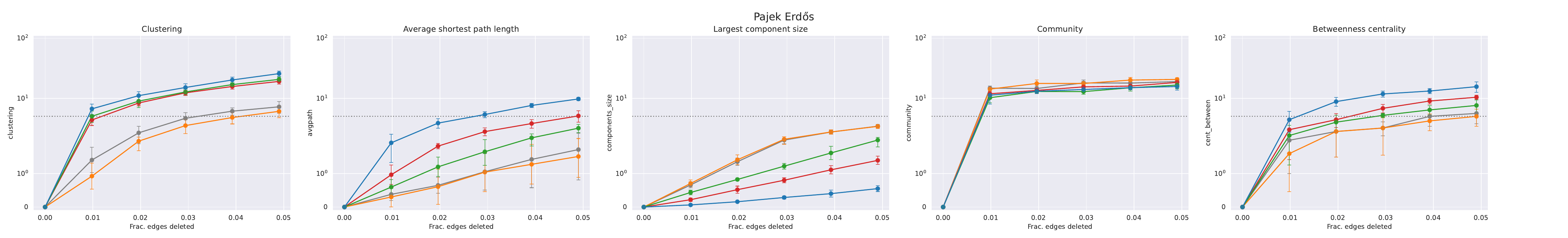}}

    \caption{Change in network properties when deleting edges. For each network (rows) and property (columns) the results show how data utility changes when deleting more edges (horizontal axis) using each of the five anonymization algorithms (color).}
	\label{fig:app:traj}
\end{center}
\end{figure}

\clearpage

\section{Runtimes}\label{app:runtimepart}
This appendix accompanies Section~\ref{sub:runtime} of the full paper.
Each figure shows both the full runtime, and time spent on the anonymization algorithm for a given network and anonymization algorithm.

\begin{figure}[htbp]
\begin{center}
    {\includegraphics[width=\textwidth]{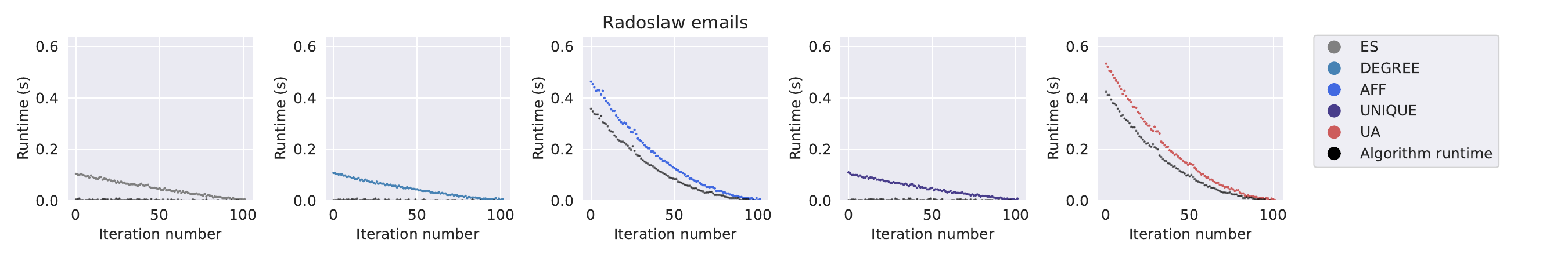}}
    {\includegraphics[width=\textwidth]{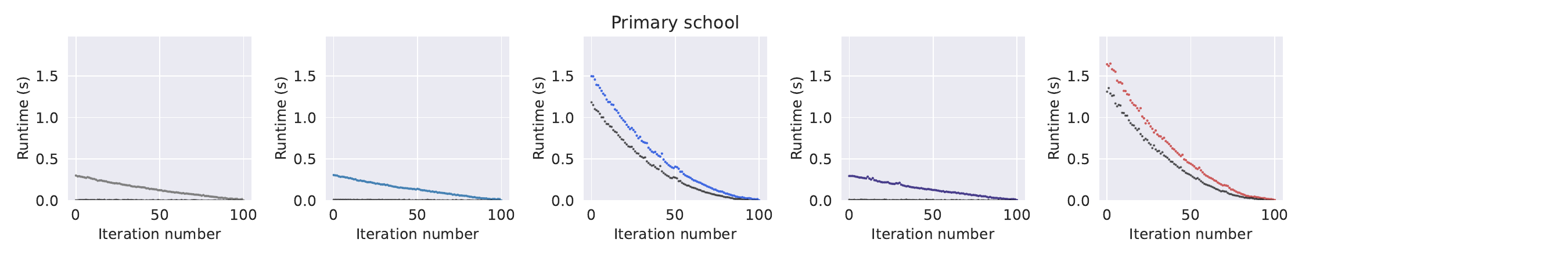}}
    {\includegraphics[width=\textwidth]{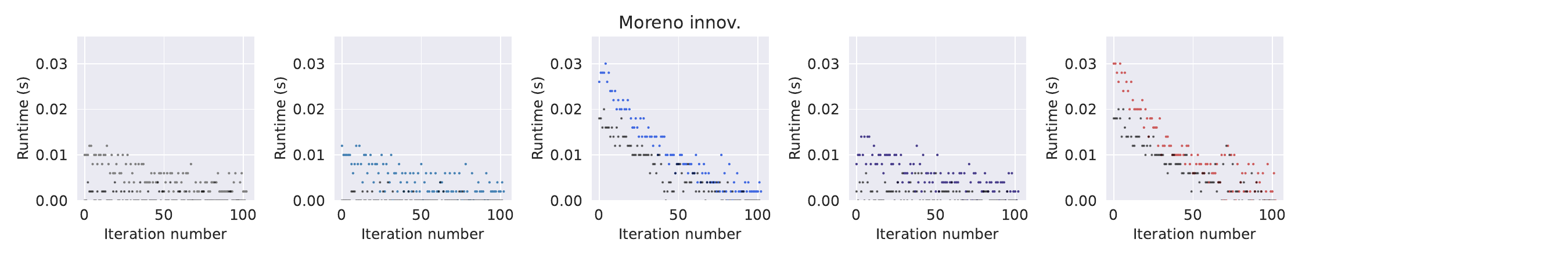}}
    {\includegraphics[width=\textwidth]{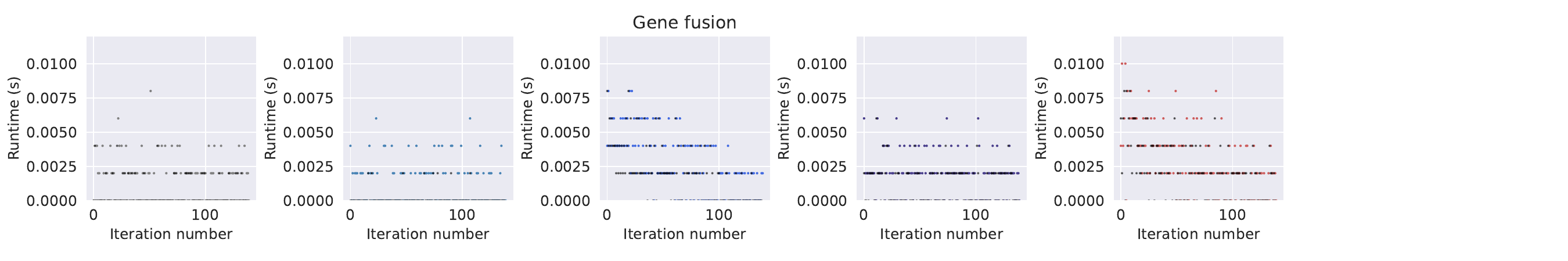}}
    {\includegraphics[width=\textwidth]{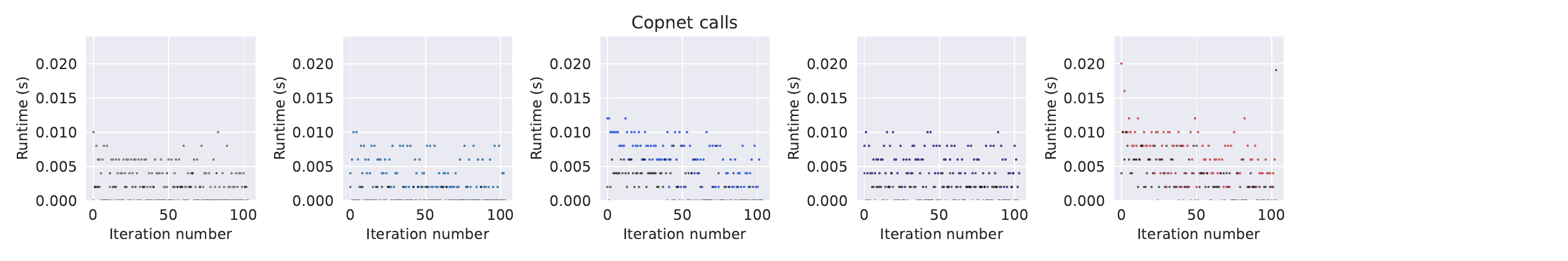}} 
    {\includegraphics[width=\textwidth]{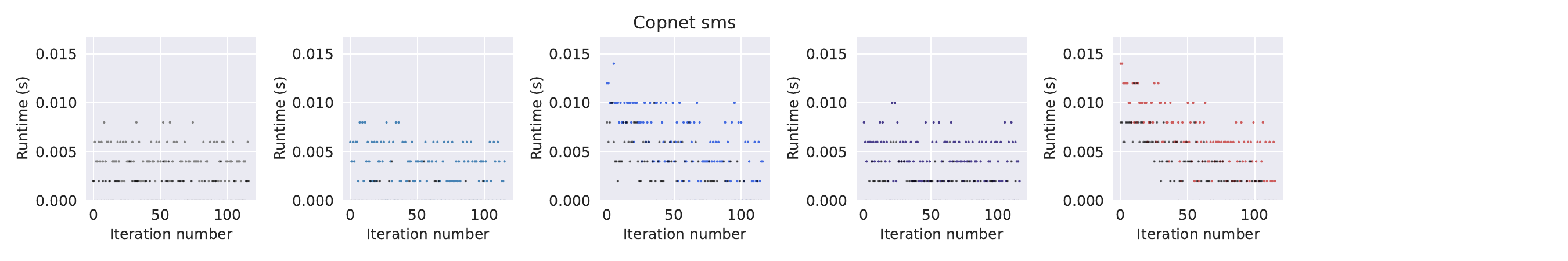}}
    \end{center}
\end{figure}

\begin{figure}[htbp]
\begin{center}   
    {\includegraphics[width=\textwidth]{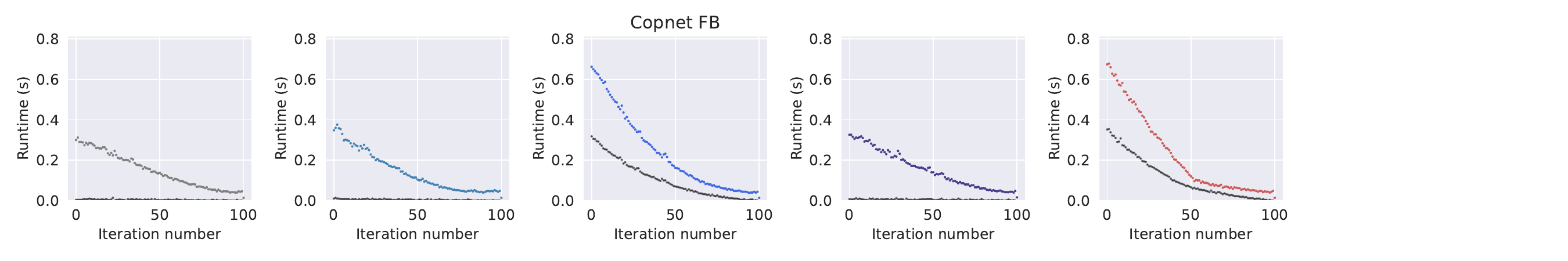}}
    {\includegraphics[width=\textwidth]{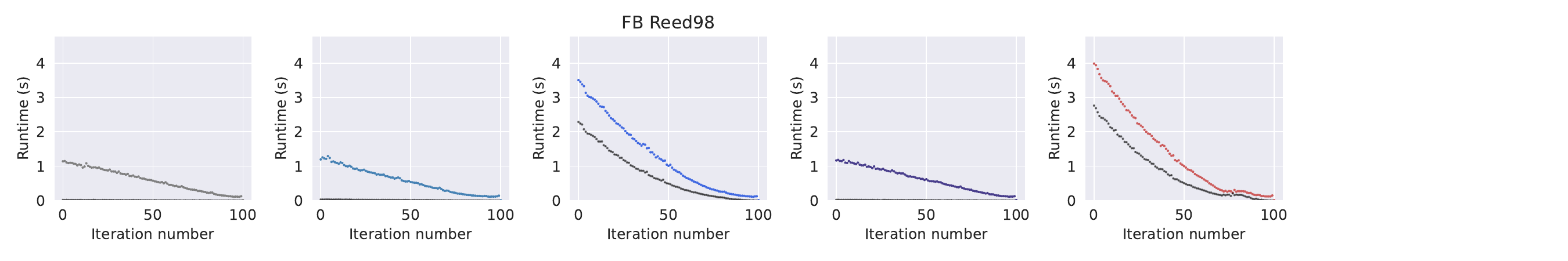}}
    {\includegraphics[width=\textwidth]{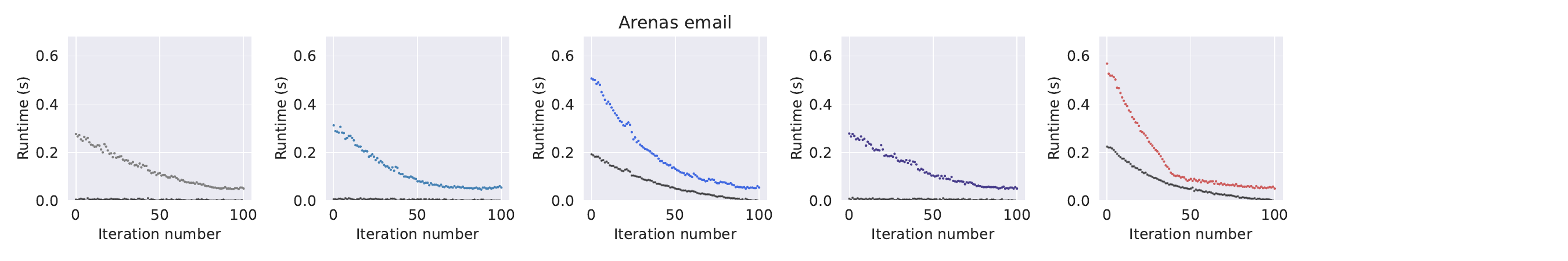}}
    {\includegraphics[width=\textwidth]{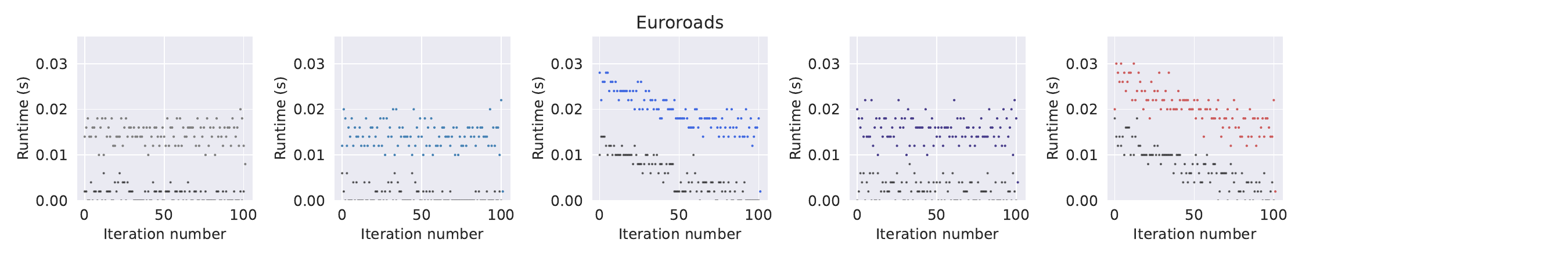}}
    {\includegraphics[width=\textwidth]{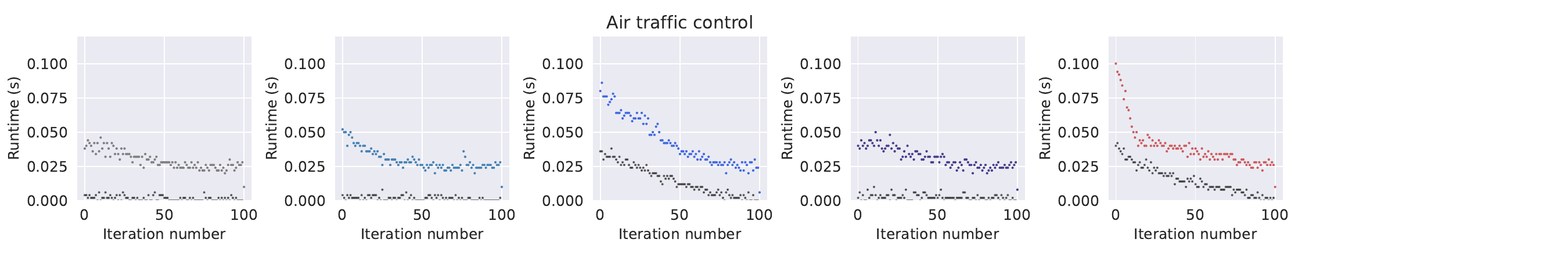}}
    {\includegraphics[width=\textwidth]{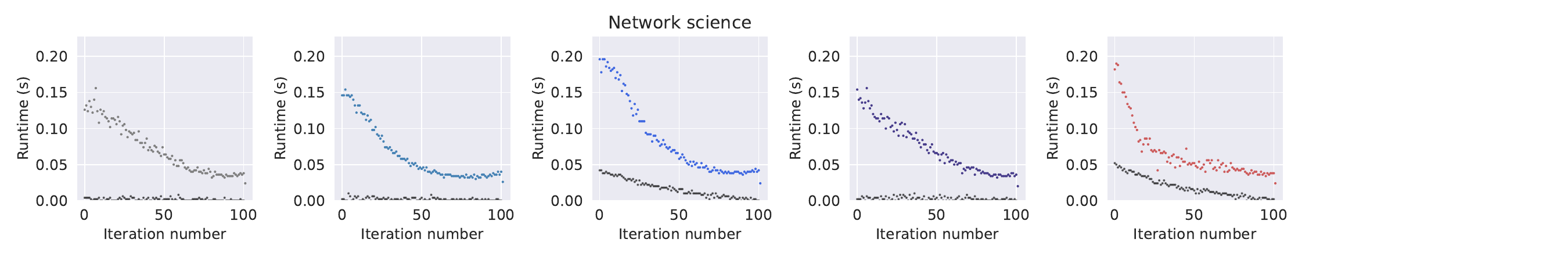}}
    {\includegraphics[width=\textwidth]{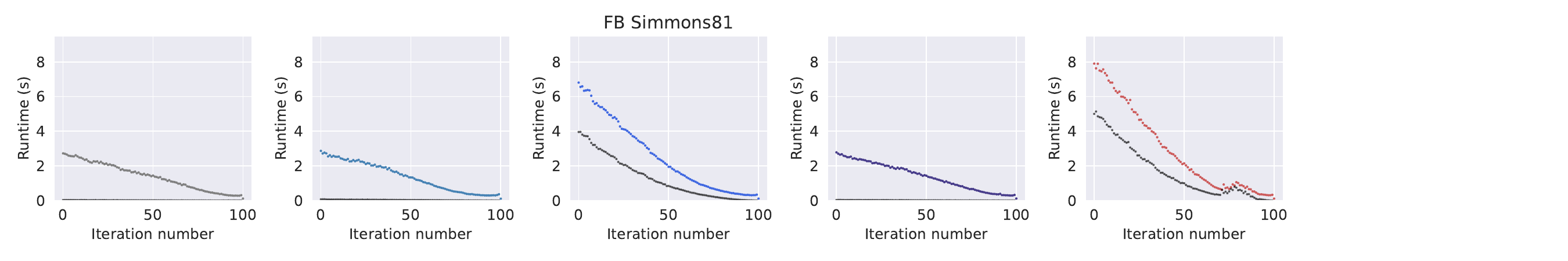}}
    {\includegraphics[width=\textwidth]{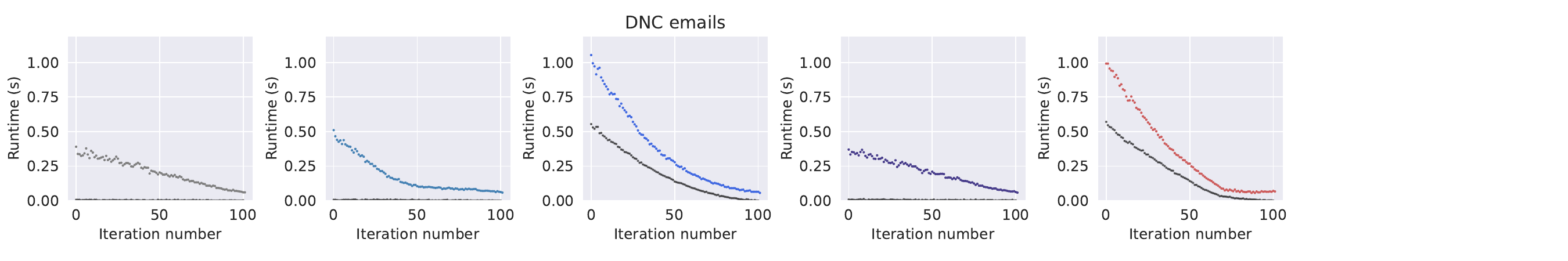}}
    {\includegraphics[width=\textwidth]{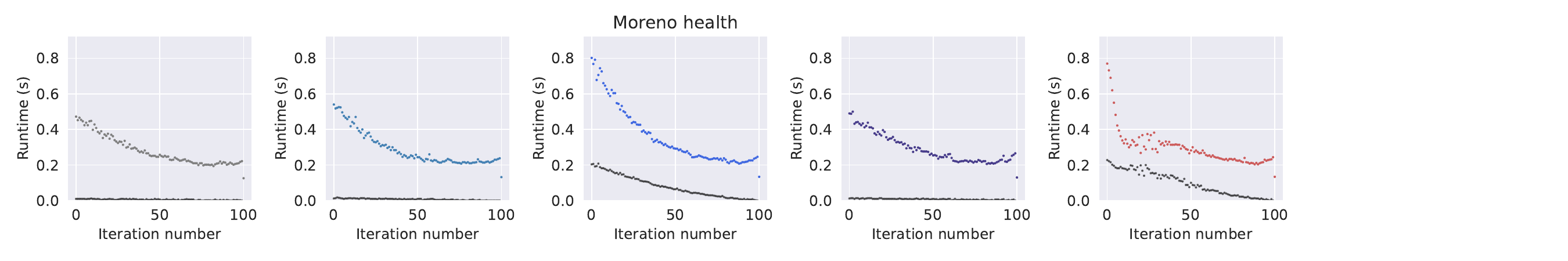}}  
\end{center}
\end{figure}

\begin{figure}[htbp]
\begin{center}
    {\includegraphics[width=\textwidth]{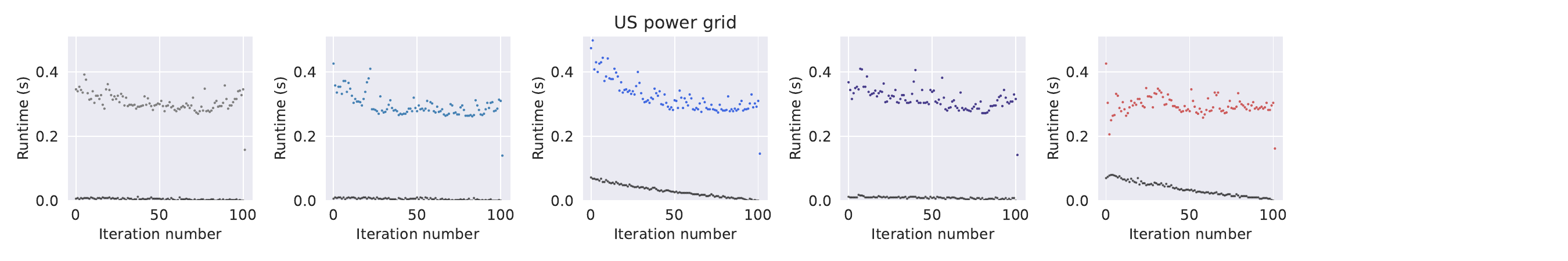}}
    {\includegraphics[width=\textwidth]{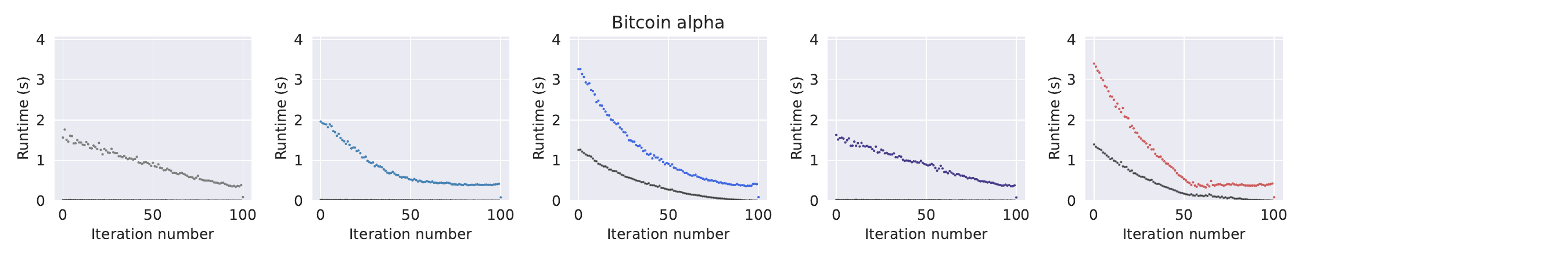}}
    {\includegraphics[width=\textwidth]{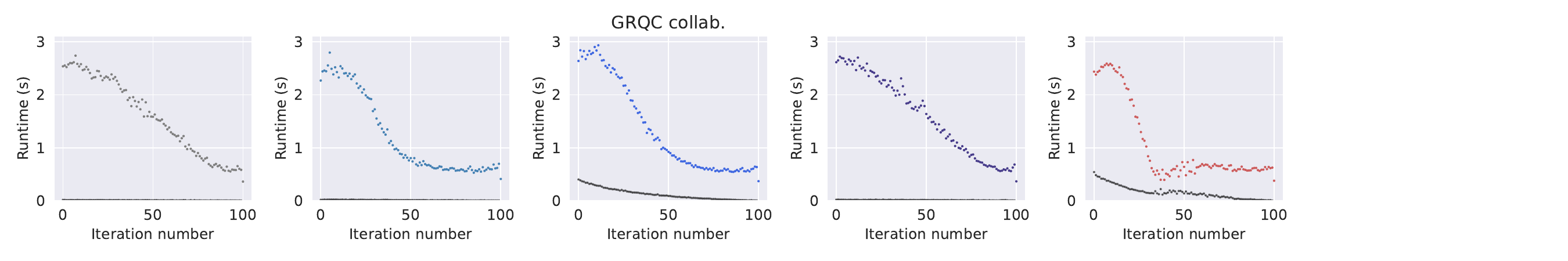}}
    {\includegraphics[width=\textwidth]{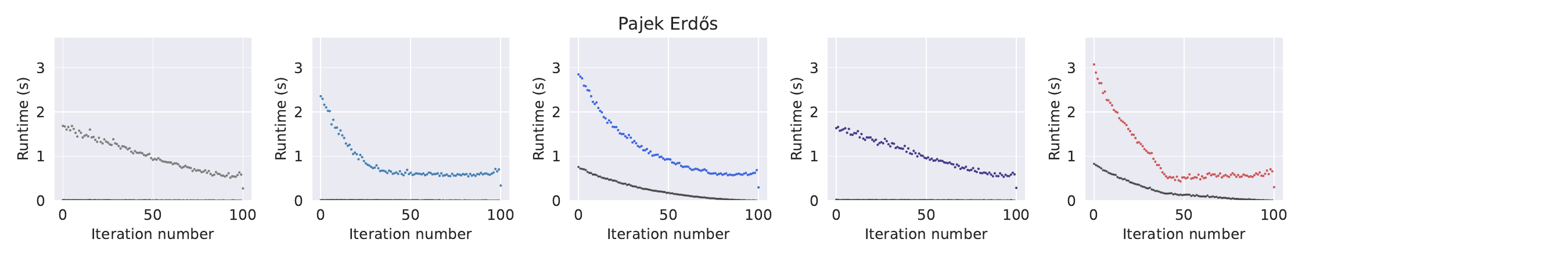}}
    {\includegraphics[width=\textwidth]{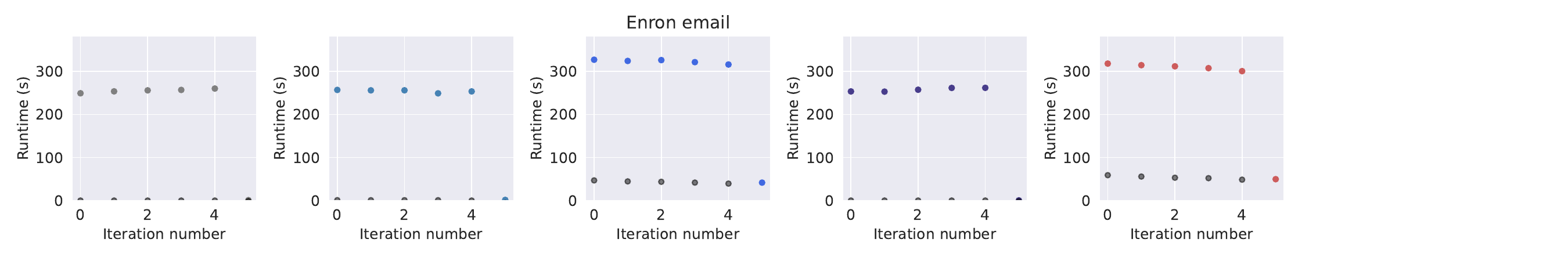}}
    {\includegraphics[width=\textwidth]{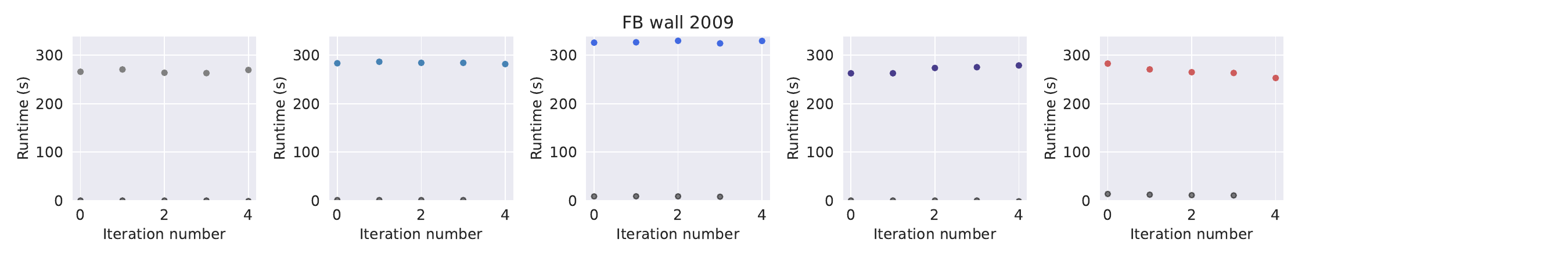}}
    {\includegraphics[width=\textwidth]{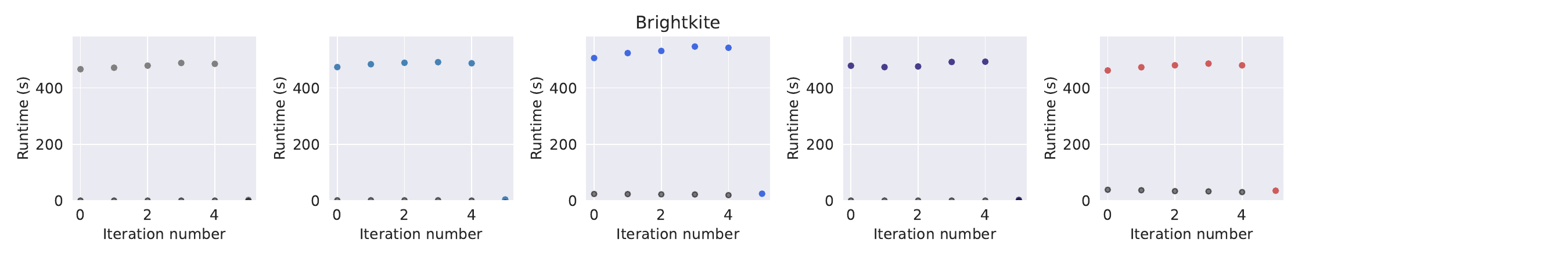}}
    \caption{Runtime per iteration. Each figure shows for each combination of network (rows) and anonymization algorithm (columns and color) the runtime per iteration. The color corresponding to the algorithm indicates the full runtime of the anonymization process, the black dots indicate the time spent on the anonymization algorithm only.}
	\label{fig:app:runtimes}
\end{center}
\end{figure}

\end{document}